\pdfoutput=1 
\documentclass[10pt,twocolumn]{article}
\PassOptionsToPackage{hyphens}{url} 

\usepackage[T1]{fontenc}
\usepackage{lmodern} 
\usepackage[margin=1in]{geometry}
\usepackage{microtype}
\usepackage{booktabs}
\usepackage{graphicx}
\usepackage{tikz}
\usetikzlibrary{positioning, arrows.meta, fit}
\usepackage[numbers,sort&compress]{natbib}
\usepackage[colorlinks=true, allcolors=blue]{hyperref}
\usepackage{amsmath, amssymb, amsthm}
\theoremstyle{definition}
\newtheorem{definition}{Definition}
\newtheorem{assumption}{Assumption}
\newtheorem{proposition}{Proposition}
\newtheorem{observation}{Observation}

\newtheorem{corollary}{Corollary}
\title{From Privacy to Workflow Integrity:\\
Communication-Graph Metadata in Autonomous Agent Interoperability}
\author{Bijaya Dangol\\ Independent Researcher\\ \texttt{dangoldbj23@gmail.com}}
\date{\today}

\begin{document}
\twocolumn[%
  \begin{@twocolumnfalse}
  \maketitle
  \begin{abstract}
Agent-interoperability protocols such as A2A and MCP standardize what agents say to one
another but assume address-based transport. Whether carried over HTTP(S) or a
content-protecting binding such as MLS-based SLIM, these transports protect message
\emph{content} yet leave the \emph{communication graph} exposed: which agent contacts
which, when, and how often. In agent systems this graph is more
consequential than a privacy framing suggests. Because endpoints are often capability-labeled,
workflows are structured and chained, and interactions are coupled to real actions, an observer
recovers more than a record of past contacts. It can recognize a \emph{recurring pending
workflow} from its opening, the task being assembled and the action likely to follow, and, since
these workflows run at machine speed, act on it before the workflow completes. The threat is
therefore one of \emph{workflow integrity}, not privacy alone.

We give a threat model for the agent communication graph and identify what makes its metadata
distinctively consequential: not stronger fingerprinting, which we measure to be comparable to
other structured machine traffic, but exposure across independent trust domains coupled to
autonomous action. We define transport- and bootstrap-layer privacy properties, evaluate
candidate transports against them, and give an A2A case study in which a metadata-protecting
binding is expressible yet surfaces the protocol's implicit identity assumptions. On a
corpus of real multi-agent A2A traffic, on traffic measured over a live A2A binding, and on a
generative model used as a controlled instrument, a classifier recovers an interaction's task
class from passive metadata well above chance ($6\times$ on the real-agent corpus), and from
only its opening. A defense-aware adversary does not overturn this, and only the full set
of properties drives recovery toward chance, at a bandwidth cost we quantify. A crawl of the
deployed ecosystem finds agent endpoints concentrated behind a few providers ($75\%$ in three),
so the observer's vantage is not hypothetical. Moving from inference to action, a metadata-only adversary in a
live testbed front-runs a competing action from a workflow's opening, winning nearly every
contested round; on the labeled corpus the same selection leverage is a capture ratio of
$0.63$ ($0.41$ from the opening fifth), which the metadata-minimization property drives to the
blind baseline. This leverage is distinct from recoverability: it is governed by the
adversary's precision on its top-ranked workflows rather than its overall accuracy, so the
integrity and privacy objectives can come apart under defense.
  \end{abstract}
  \bigskip
  \end{@twocolumnfalse}
]

\section{Introduction}\label{sec:intro}
AI agents built by different vendors are increasingly made to interoperate through
open protocols. A2A~\citep{a2a}, now hosted by the Linux Foundation, and MCP~\citep{mcp} let agents
discover one another, delegate tasks, and increasingly transact on behalf of users and
organizations. These protocols standardize the \emph{content} and structure of agent
messages, but assume a conventional, address-based transport: agents are reachable
at URLs or stable names, and messages travel predominantly over HTTP(S).

Transport security here has focused, reasonably, on protecting message content: TLS
in transit and a growing set of end-to-end schemes that keep payloads from
intermediaries. What this leaves untouched is the \emph{communication graph}: the
record of which agent contacts which, when, how often, and how much data flows.
Because routing requires addressing and addresses are identifiers, this graph is
visible to network observers, relays, and registries even when every payload is
encrypted.

The graph is more than a privacy concern. Endpoints are often capability-labeled,
workflows are structured and chained, and interactions are coupled to real actions.
From such a graph an observer reads the \emph{shape of a task in progress}, not merely
a record of past contacts, and at machine speed can act on that shape before the
workflow completes. What is exposed is then a matter of \emph{integrity}, not privacy
alone: the observer holds predictive leverage over actions that have not yet occurred.
Existing agent-protocol threat models examine authentication, identity, and payload
leakage; the communication graph, with its prospective, action-coupled character, has
drawn little attention. This paper develops it systematically.

Our contributions are:
\begin{enumerate}
  \item A \textbf{threat model} for the agent-interop \emph{communication graph} as
  a metadata surface, separate from payload confidentiality
  (\S\ref{sec:threat}, \S\ref{sec:problem}).
  \item An account of \textbf{why agent metadata is distinctively consequential}
  (semanticity, prospectivity, vantage, and actuation), which locates the novelty not
  in inference strength, which we measure to be comparable to non-agent traffic, but in
  cross-trust-domain exposure and machine-speed actuation, reframing the threat from
  privacy to the \emph{integrity of autonomous workflows} (\S\ref{sec:different}).
  \item A \textbf{transport- and bootstrap-privacy property framework} (unlinkability,
  no central observer, deniability, metadata minimization, and discovery privacy) and
  an evaluation of candidate transports against it (\S\ref{sec:properties},
  \S\ref{sec:transports}).
  \item An \textbf{A2A case study} showing a metadata-protecting binding is
  expressible but surfaces the protocol's implicit identity assumptions, and a
  \textbf{reconciliation} with the ecosystem's identity and reputation direction
  (\S\ref{sec:casestudy}, \S\ref{sec:discussion}).
  \item An \textbf{empirical evaluation} whose headline evidence is a \textbf{corpus of real
  official-SDK multi-agent A2A traffic} and traffic \textbf{measured over a live A2A binding},
  with a generative model as a \emph{controlled instrument} for parameter sweeps and the
  actuation analytics: a label-blind network observer recovers task class at $6\times$ chance
  and from only a short prefix of a workflow; only the full set of properties collapses this
  recovery toward chance, and a \textbf{defense-aware adversary} trained on the protected
  traffic does not undo it; we report the \textbf{bandwidth cost} of each point on the
  defense frontier and how leakage scales with \textbf{partial adoption}, and the recovery is
  \textbf{composition-specific}, vanishing on workflow shapes never seen in training. A comparison
  against \textbf{production microservice traffic} finds recovery at least as high there,
  locating the contribution in vantage and actuation, not inference strength; and a \textbf{crawl}
  of the deployed MCP ecosystem finds agent endpoints concentrated behind a few providers
  ($75\%$ in three), grounding the vantage axis. Moving from inference to action,
  the recovered signal carries decision-theoretic \textbf{leverage}, which we
  \textbf{demonstrate as a live front-running suite}: a metadata-only adversary, from a
  workflow's opening, races and wins a competing action across three attack scenarios,
  and the metadata-minimization property collapses every one to the blind baseline
  (\S\ref{sec:eval}).
\end{enumerate}

The contribution is not a new attack. The inference is standard
traffic analysis, no stronger on agent traffic than on production microservice traffic
(\S\ref{subsec:distinct}), and the defenses are taken unchanged from the anonymous-communication
and website-fingerprinting-defense literature~\citep{tamaraw, wtfpad, front, surakav}. What is
new is where this familiar machinery bites: a readable agent graph exposed \emph{across
independent trust domains} and coupled to machine-speed action (\S\ref{sec:different}), an
exposure an ecosystem crawl shows is real rather than hypothetical (\S\ref{sec:threat}); and a
\emph{measure} of that action, the capture ratio $\kappa$ (\S\ref{subsec:actuation}), distinct
from recoverability and governed by top-ranked precision rather than overall accuracy
(Proposition~\ref{prop:kappa}), so the integrity and privacy objectives come apart under defense.
Neither website fingerprinting, which classifies a \emph{completed} trace, nor front-running,
which needs a content-visible mempool, occupies this regime (Table~\ref{tab:position}).

\section{Background}\label{sec:background}

\subsection{Agent interoperability protocols}
A2A models interoperation as \emph{tasks} exchanged between a client and a remote
agent. Agents publish \emph{Agent Cards} (metadata documents at well-known URLs
that declare capabilities, endpoints, and authentication) and communicate over one
of several \emph{bindings} (JSON-RPC, gRPC, or HTTP+JSON), all over
HTTPS~\citep{a2a}. Operations are asynchronous: a call returns immediately, and task
updates arrive by polling, server-sent streaming, or push notifications to a
client-provided webhook~\citep[\S3.1.7]{a2a}. A2A also admits \emph{custom protocol
bindings} for transports beyond the core set~\citep[\S5]{a2a}, the extension point
we use in \S\ref{sec:casestudy}. MCP plays a complementary
(agent-to-tool) role but shares the same address-based, HTTP-oriented assumptions.

\subsection{Transport security today}
Beyond TLS, recent bindings strengthen content protection: SLIM/SLIMRPC provides
broker-less delivery with MLS end-to-end encryption~\citep{slim, mls}, so that no
central intermediary reads message content. These mechanisms target confidentiality
and, for SLIM, the removal of a trusted broker; none aims at concealing the
communication graph.

\subsection{Metadata-protecting transports}
A separate lineage protects communication \emph{metadata}: mix
networks~\citep{chaum1981mix}, onion routing~\citep{tor}, mixnets~\citep{nym}, and
identity-less messaging such as SimpleX's SMP~\citep{simplex}. These were built for
human or general messaging; \S\ref{sec:transports} asks what they offer when
repurposed as agent-interop transports.

\section{System and Threat Model}\label{sec:threat}

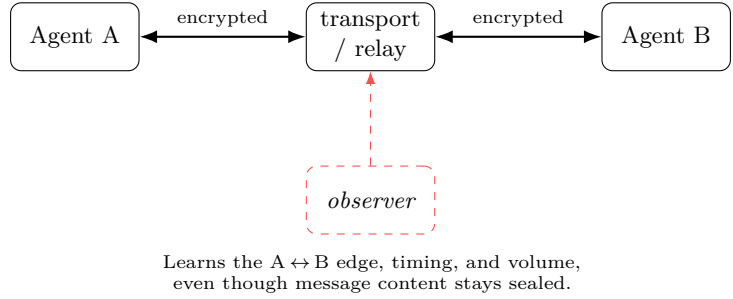
\begin{figure*}[t]
\centering
\begin{tikzpicture}[>=Latex, node distance=2.2cm,
  box/.style={draw, rounded corners, minimum width=1.7cm, minimum height=0.9cm,
              font=\small, align=center}]
  \node[box] (A) {Agent A};
  \node[box, right=of A] (R) {transport\\/ relay};
  \node[box, right=of R] (B) {Agent B};
  \draw[<->, thick] (A) -- node[above, font=\scriptsize] {encrypted} (R);
  \draw[<->, thick] (R) -- node[above, font=\scriptsize] {encrypted} (B);
  \node[box, draw=red!70, dashed, below=1.25cm of R, font=\small\itshape] (O) {observer};
  \draw[->, red!70, dashed] (O) -- (R);
  \node[below=0.15cm of O, font=\scriptsize, text width=6.4cm, align=center]
    {Learns the A\,$\leftrightarrow$\,B edge, timing, and volume, even though
     message content stays sealed.};
\end{tikzpicture}
\caption{Content encryption protects the payload but not the communication graph.
An observer at the network or at an intermediary learns who talks to whom, when, and
how often; in agent systems, often capability-labeled endpoints and their sequence
further reveal the task in progress (\S\ref{sec:different}).}
\label{fig:graph}
\end{figure*}

\subsection{System model}
We consider a set of agents $\mathcal{A}=\{a_1,a_2,\dots\}$ that interoperate by
exchanging messages under an interop protocol such as A2A or MCP. Two agents
communicate over a \emph{transport binding} that realizes the protocol's abstract
operations (request/response, streaming updates, and notifications) over a
concrete transport. A binding may route through one or more \emph{intermediaries}
(relays, gateways, or brokers), and the protocol may use a \emph{registry} for
capability discovery and connection bootstrap.

An \emph{interaction} between $a_i$ and $a_j$ is the set of messages exchanged to
complete one logical exchange (in A2A, the lifecycle of a task). Each message $m$
carries a transport-visible descriptor
\[
  \mathrm{obs}(m) = (\mathrm{src},\,\mathrm{dst},\,t,\,\ell,\,d),
\]
its endpoint identifiers, timestamp, length, and direction. Notably
$\mathrm{obs}(m)$ excludes the message \emph{content}, which we assume encrypted
(Assumption~\ref{asm:content}).

\subsection{The communication graph}
Over a period of operation, interactions induce a \emph{communication graph}
$G=(V,E)$: $V$ is the set of transport-visible agent identifiers, and each edge
$e\in E$ records that two endpoints interacted, annotated with timing, frequency,
and volume. A \emph{linkage} relation maps transport-visible identifiers to
persistent agent or operator identities. The assets we protect are $G$, this
linkage, and, as \S\ref{sec:different} argues, the predictive leverage over future
action that $G$ confers; not the content, which is protected by other means.

\subsection{Adversary model}
We model \emph{honest-but-curious} adversaries distinguished by vantage point
(Table~\ref{tab:adv}); the threat is that $G$ leaks without any active attack. A
network observer $\mathcal{N}$ sees $\mathrm{obs}(m)$ for messages on observed
links; an intermediary $\mathcal{R}$ sees what it forwards; a registry
$\mathcal{G}$ sees discovery lookups and connection bootstrap; a participating or
log-retaining endpoint $\mathcal{E}$ sees its own interactions, and more under
collusion. Adversaries may collude to widen their coverage of $E$. Our leakage analysis needs
only passive observation: the objective ranges from reconstructing $G$ to inferring
the pending workflow it encodes (\S\ref{sec:different}). \emph{Acting} on that
inference (to preempt or interfere) may require separate active capabilities,
which only strengthen the adversary.

That a few vantage points already cover much of $E$ is not hypothetical. Crawling the
public Model Context Protocol registry for every remotely reachable agent endpoint and
resolving each to its hosting autonomous system, we find $1{,}424$ distinct endpoints
concentrated heavily: the single largest provider (Cloudflare) fronts $42\%$, the top
three (Cloudflare, Amazon, Google) $75\%$, and the top ten $92\%$. This measures hosting
and CDN concentration, not graph omniscience: a provider that fronts a server sees the
\emph{inbound} edges to that server (and, terminating TLS, their $\mathrm{obs}(m)$), not the
full cross-domain graph. But a party fronting $42\%$ of reachable endpoints observes a large
fraction of all edges without being global, which is exactly the coverage
Definition~\ref{def:nocentral} rules out: a network or intermediary adversary need not see
everything to reconstruct much of $G$, and in the deployed ecosystem a few clouds are
already positioned to. The registry is a snapshot of a fast-moving, partly experimental
ecosystem, so we read the concentration as indicative rather than precise. The
\emph{no central observer} property is one the current infrastructure does not provide.

\begin{table*}[t]
\centering\small
\begin{tabular}{@{}lll@{}}
\toprule
Adversary & Vantage point & Observes \\
\midrule
$\mathcal{N}$ network & links / paths & $\mathrm{obs}(m)$: endpoints, timing, volume \\
$\mathcal{R}$ intermediary & relay / gateway / broker & forwarded src/dst, timing, volume \\
$\mathcal{G}$ registry & discovery service & lookups, connection bootstrap \\
$\mathcal{E}$ endpoint / log & a participant or its logs & own interactions; more under collusion \\
\bottomrule
\end{tabular}
\caption{Adversary classes. All are passive (honest-but-curious) in the base
model and may collude to increase coverage of $E$.}
\label{tab:adv}
\end{table*}

\subsection{Trust assumptions and scope}
\begin{assumption}[Content confidentiality]\label{asm:content}
Message content is end-to-end encrypted under secure primitives; the adversary
learns nothing from payloads.
\end{assumption}
We deliberately grant the \emph{strongest} content protection so as to isolate the
metadata axis: any leakage we identify is leakage that content encryption, however
strong, does not prevent.
\begin{assumption}[No trusted graph custodian]\label{asm:nocustodian}
No single party is trusted to observe the full communication graph.
\end{assumption}
Out of scope are content confidentiality and payload data minimization (assumed
handled elsewhere), application-level authorization, and side channels outside the
transport.

\section{The Communication-Graph Metadata Problem}\label{sec:problem}

\subsection{Content and metadata are independent}
The starting point is a simple but consequential observation: content
confidentiality and communication-graph privacy are \emph{independent}, because
they protect different things. An interaction can be perfectly content-confidential
and still be fully graph-exposed.

Concretely, under Assumption~\ref{asm:content} the payload reveals nothing, yet
routing still requires the transport to address the destination. When endpoints
are named by persistent identifiers, those identifiers appear as $\mathrm{src}$
and $\mathrm{dst}$ in every $\mathrm{obs}(m)$ and directly reveal the edge; the
timing $t$, length $\ell$, and direction $d$ are visible regardless of encryption.
Since $G$ is by definition the set of such edges, any adversary that observes them
reconstructs $G$ no matter how strong the content protection is.

We state this as an observation, not a theorem: it is close to
definitional. The contribution is not the claim that metadata leaks (that is well
understood for communication systems in general) but a systematic account of
\emph{which} metadata leaks in agent-interop protocols, to \emph{whom}
(\S\ref{sec:threat}), what a transport must provide to prevent it
(\S\ref{sec:properties}), and the protocol-level consequences of providing it
(\S\ref{sec:casestudy}).

\subsection{A walk through the A2A task lifecycle}
Let a client $a_c$ delegate a task to a server $a_s$ under A2A, with all content
encrypted. At \emph{discovery}, $\mathcal{G}$ observes that $a_c$ resolved $a_s$.
At \emph{connection setup}, $\mathcal{N}$ and any on-path $\mathcal{R}$ observe the
$a_c\!\leftrightarrow\!a_s$ edge. On \emph{message/send} they observe timing and
size; across \emph{streaming or polled updates} they observe cadence and volume; a
\emph{push notification} additionally exposes $a_c$'s callback endpoint. By
\emph{completion}, although no adversary has read a single field of the task,
$\mathcal{N}$, $\mathcal{R}$, and $\mathcal{G}$ jointly learn that $a_c$ engaged
$a_s$, when, how often, and how much data flowed, and across many tasks the
shape of the agents' relationships.

\subsection{Why current bindings do not address it}
The A2A bindings over HTTPS (JSON-RPC, gRPC, HTTP+JSON) protect content with TLS
but address agents by URL, so $\mathcal{N}$ and $\mathcal{R}$ obtain the edge
directly. SLIM/SLIMRPC removes the central content-reading broker via MLS, yet
routes by a persistent structured name; $\mathcal{R}$ and $\mathcal{N}$ still
obtain the edge, and the persistent name supplies the linkage of
\S\ref{sec:threat}. None of these target $G$: they protect content, which is
independent of the graph.

\subsection{Problem statement}
\begin{definition}[Metadata-protecting binding]\label{def:mpb}
A transport binding is \emph{metadata-protecting} against an adversary class if,
from that adversary's observations, it cannot reconstruct the communication graph
$G$, nor infer the pending workflow it encodes (\S\ref{sec:different}), beyond a
bounded, unlinkable view, as made precise by the properties of \S\ref{sec:properties}.
\end{definition}

\section{Why Agent Metadata Is Different}\label{sec:different}
Generic communication metadata reveals \emph{that} parties communicated. In agent
interoperability the same graph reveals a \emph{task in progress}, along four axes:
semanticity, prospectivity, vantage, and actuation. Not all are new to the agent setting.
Semanticity and prospectivity make the graph \emph{readable}, and we will show they read
strongly (\S\ref{sec:eval}); but they are not unique to agents. Structured
machine-to-machine traffic in general carries them, and a measurement against
production microservice traffic (\S\ref{subsec:distinct}) finds task-class recovery
there at least as high. What distinguishes the agent setting is not that its
metadata reveals more, but \emph{who} is positioned to read it and \emph{what acting
on it does}, the vantage and actuation axes we develop below. The contribution is the
reframing those two force, not a claim that agent traffic fingerprints better.

\paragraph{Semanticity.} Agent endpoints, tools, and registry entries are often
semantically meaningful rather than opaque addresses. Agent Cards advertise skills, registries are queried by capability, and
MCP tools are named by function. Observing that a client contacted a
``contract-review'' agent or invoked a ``payments'' tool reveals the \emph{class} of
task, not merely that an interaction occurred. This is the explicit-label analogue
of website-fingerprinting attacks, where the class of activity is inferred from
encrypted-traffic metadata~\citep{wfp}, except that here the label is advertised
rather than inferred. What an observer actually recovers depends on vantage point
and binding: a registry sees capability queries directly, whereas a pure network
observer may recover labels only indirectly, through discovery lookups, Agent Card
fetches, structured names, or repeated endpoint patterns.

\paragraph{Prospectivity.} Agent workflows are structured and chained: discovery
precedes delegation, delegation precedes tool invocation, and updates follow. Early
steps can therefore predict later ones. An observer who recognizes the opening of a
familiar workflow may anticipate its trajectory before it completes, rather than
learning of it only afterward.

\paragraph{Vantage.} Semanticity and prospectivity are also present in, say, a
service mesh, but there the only party positioned to read the graph is the operator
who runs the cluster and already sees everything inside it; the exposure is internal
and the reader is already trusted. Agent interoperation inverts this. Discovery is
cross-organizational by design, agents reach one another at public addresses across
independent trust domains, and a principal's authority is delegated over those links.
The same readable graph is therefore exposed to parties who are \emph{not} otherwise
privy to the workflow, network observers, relays, and shared registries between
mutually distrusting organizations, rather than to a single operator who already holds
the data. The novelty is in the topology of exposure, not the strength of the signal.

\paragraph{Actuation.} Agent interactions often \emph{trigger actions} directly,
without a human reviewing each step. The graph is thus coupled to consequences in
the world: influencing or interrupting the observed workflow can change what the
agents actually do. This is the axis on which the agent setting departs furthest from
prior traffic analysis. Recovering the class of an internal microservice call tells an
external party little it can act on; recovering the pending shape of a cross-domain
agent workflow, early and at machine speed, yields \emph{leverage} over an action that
has not yet happened. \S\ref{subsec:actuation} measures this leverage directly, as the
quantity that recoverability alone does not capture.

As an illustration, a lookup for a sanctions-screening agent, followed by
payment-settlement and contract-review calls, suggests a cross-border transaction
being assembled, revealing the \emph{kind} of deal in progress well before it
completes, without a single payload being read.

Together these shift the adversary's objective. From passive observation alone a
graph observer may infer historical relationships and, beyond them, \emph{pending
intent and workflow trajectory}. The \emph{harm} comes when the adversary acts on
that inference through separate, active channels: poisoning discovery, preempting a
negotiation, triggering a competing action. Because the workflow is structured and
runs at machine speed, such a move can land before the workflow completes. The
pattern is familiar from front-running in decentralized exchanges, where adversaries
watch pending-transaction metadata and act ahead of it~\citep{flashboys}, a regime
that exists not because trading is new but because execution became automated; agent
actuation stands to human service-use as that front-running stands to human trading.
The risk arises wherever workflows are capability-labeled and structured, exposed
across trust boundaries, and potentially across many application domains rather than
one.

The framing shifts accordingly. Protecting the communication graph is not merely a
privacy question, concealing who interacts; it concerns the \emph{integrity and contestability
of autonomous workflows}: their freedom to execute, and to be steered by their
principals rather than by an outside observer who holds predictive leverage over
machine-speed action. Section~\ref{sec:eval} measures how much intent the graph
leaks, inferring task class from endpoint and sequence metadata, how that recovery
compares to non-agent traffic, and what acting on it is worth; here the threat
serves as the design motivation. A scope note: what becomes transport-visible is
\emph{inter-agent and inter-tool} coordination, not an agent's internal, local
planning.

Table~\ref{tab:position} places the resulting threat against the two literatures it is most
likely to be assimilated to. Website fingerprinting and microservice traffic analysis recover
a class but read a \emph{completed} interaction, and the latter is read by the operator who
already owns the cluster; decentralized-exchange front-running acts \emph{ahead} of a pending
action, but needs a content-visible mempool and structural ordering power. The agent setting
is the only one that combines all of: an adversary that \emph{acts} rather than merely
recovers, \emph{before} the workflow completes, \emph{across} independent trust domains, from
\emph{content-encrypted} metadata alone. This is the cell our contribution occupies, and
\S\ref{subsec:actuation}--\S\ref{subsec:demo} measure and then demonstrate it.

\begin{table*}[t]
\centering\footnotesize
\begin{tabular}{@{}lllll@{}}
\toprule
 & Adversary acts on & When & Trust boundary & Needs content \\
\midrule
Website fingerprinting~\citep{wfp, df} & recoverability & after completion & single domain & no \\
Microservice traffic analysis~\citep{oriondep} & recoverability & after the fact & intra-domain & no \\
DEX front-running / MEV~\citep{flashboys} & actuation & before execution & open mempool & yes (mempool) \\
\textbf{This work} & \textbf{actuation} & \textbf{before completion} & \textbf{cross-domain} & \textbf{no} \\
\bottomrule
\end{tabular}
\caption{Positioning against the nearest prior work. Each component is individually familiar;
the agent-interop setting is distinguished by occupying the combination, acting on an
\emph{incomplete}, cross-domain workflow from content-encrypted metadata. The novelty is in
this combination and in measuring it (\S\ref{subsec:actuation}), not in the inference, which
is standard traffic analysis (\S\ref{subsec:distinct}).}
\label{tab:position}
\end{table*}

\section{Privacy Properties for Transport and Bootstrap}\label{sec:properties}
The following properties span transport and bootstrap and are protocol-independent;
for each we note the adversary capability it removes.

\begin{definition}[Unlinkability]\label{def:unlink}
An adversary cannot tell whether two observed interactions involve the same agent,
nor link a transport-visible identifier to a persistent agent identity. This
requires that identifiers not be stable across interactions: each interaction uses
a fresh identifier unlinkable to the agent's others.
\end{definition}
\noindent Identifier freshness is the mechanism; it denies the edge-linkage of
\S\ref{sec:threat} to $\mathcal{N}$ and $\mathcal{R}$, and the persistent-identity
linkage to all classes.

\begin{definition}[No central observer]\label{def:nocentral}
No single adversary vantage point observes more than a small fraction of $E$;
reconstructing $G$ requires collusion among multiple independent parties.
\end{definition}
\noindent This targets the global view of a network observer $\mathcal{N}$ or a
shared intermediary $\mathcal{R}$, and follows Assumption~\ref{asm:nocustodian}.

\begin{definition}[Deniability]\label{def:deny}
An interaction leaves no transferable transcript that cryptographically binds a
specific agent to participation; any party can plausibly deny it.
\end{definition}
\noindent This targets a logging or colluding endpoint $\mathcal{E}$.

\begin{definition}[Metadata minimization]\label{def:minimize}
The observable descriptors $(t,\ell,d)$ are reduced (e.g., padded, batched, or
mixed) so that timing and volume do not distinguish interactions.
\end{definition}
\noindent This targets traffic analysis by $\mathcal{N}$ and $\mathcal{R}$ that
survives even fresh identifiers.

\begin{definition}[Discovery privacy]\label{def:discovery}
Capability lookup and connection bootstrap do not reveal the requested capability,
the selected peer, or the resulting interaction edge to an untrusted registry or
transport intermediary.
\end{definition}
\noindent This targets the registry $\mathcal{G}$ and the early, pre-interaction
leakage that \S\ref{sec:different} identifies as especially sensitive.

Definitions~\ref{def:unlink}--\ref{def:minimize} are wire-transport properties;
\S\ref{sec:transports} evaluates how far real transports meet them. Discovery
privacy (Definition~\ref{def:discovery}) is realized at the bootstrap layer instead,
and \S\ref{sec:casestudy} shows how an identity-less binding can provide it through
out-of-band exchange. Together they make a binding metadata-protecting
(Definition~\ref{def:mpb}) against the corresponding adversaries; in the terms of
\S\ref{sec:different} they bound an adversary's predictive leverage by denying the
identity, timing, and discovery cues that make workflow inference possible, so they
protect the integrity of autonomous workflows, not only privacy.

What the properties buy, and what each subset leaves, is sharpest as an
indistinguishability game in the style standard for privacy notions. A binding $\Pi$
realizing some subset of the properties induces, for a workflow $w$, a distribution
$\mathsf{View}_\Pi(w)$ over network-view transcripts $(\mathrm{obs}(m))_m$, where each
$\mathrm{obs}(m)=(\iota_m,t_m,\ell_m,d_m)$ records an identifier, a timestamp, a length,
and a direction. Write $N(w)=(|w|,(d_m)_m)$ for the message-count-and-direction vector,
the structural channel that no per-message reshaping touches.

\begin{definition}[Network-view indistinguishability]\label{def:indgame}
In experiment $\mathsf{Ind}_{\mathcal{A},\Pi}$ the adversary $\mathcal{A}$ (a network
observer $\mathcal{N}$, or a relay $\mathcal{R}$, which sees the same $\mathrm{obs}$ on
forwarded links) names two workflows $w_0,w_1$; the challenger draws $b\in\{0,1\}$
uniformly, runs $w_b$ over $\Pi$, and returns a transcript $\tau\sim\mathsf{View}_\Pi(w_b)$;
$\mathcal{A}$ outputs $b'$ and wins if $b'=b$. The advantage is
$\mathsf{Adv}_{\mathcal{A},\Pi}=\lvert 2\Pr[b'=b]-1\rvert$, and $\Pi$ is network-view
indistinguishable if $\mathsf{Adv}_{\mathcal{A},\Pi}=0$ for every $\mathcal{A}$ and every
pair $w_0,w_1$. The notion is information-theoretic and passive: content is already
end-to-end encrypted (Assumption~\ref{asm:content}), so the only signal in play is
$\mathrm{obs}$, and a $K$-class recovery adversary's advantage over the $1/K$ prior is at
most $\max_{w_0,w_1}\mathsf{Adv}_{\mathcal{A},\Pi}$ by a routing argument over class pairs.
\end{definition}

\begin{observation}[Each property closes one channel; only the full set closes all]\label{thm:channels}
Against the network adversary, the wire properties act on disjoint channels of
$\mathsf{View}_\Pi(w)$:
\begin{enumerate}
\item unlinkability (Def.~\ref{def:unlink}) makes the view invariant under identifier
renaming, so the identifier marginal $(\iota_m)_m$ is a fixed distribution independent of
$w$;
\item metadata minimization (Def.~\ref{def:minimize}) fixes every $\ell_m$ to a constant
cell and every $t_m$ to a deterministic function of the message index, so the size and
timing marginals depend on $w$ only through $N(w)$;
\item constant-rate, fixed-length, full-duplex cover makes $N(w)$ itself constant.
\end{enumerate}
Under all three, $\mathsf{View}_\Pi(w)$ is the same distribution for every $w$, hence
$\mathsf{Adv}_{\mathcal{A},\Pi}=0$ for all $\mathcal{A}$: $\Pi$ is network-view
indistinguishable. Any proper subset leaves at least one channel open, and then some
adversary achieves
$\mathsf{Adv}_{\mathcal{A},\Pi}=\mathrm{SD}\!\left(\mathsf{View}_\Pi(w_0),\mathsf{View}_\Pi(w_1)\right)$,
the statistical distance the open channel induces between a worst-case $w_0,w_1$, which is
positive; no proper subset is network-view indistinguishable.
\end{observation}
\noindent The argument is immediate from the definitions: each property equalizes the named
marginal across all $w$, and the advantage of an optimal distinguisher equals the
statistical distance between the two views, which factors through whatever channels remain
trace-dependent. Closing all of them leaves identical views and zero advantage; leaving one
open leaves a pair it separates. So ``only as a set'' is not a heuristic but a statement
about coverage: a defense that omits a channel is bounded away from indistinguishability by
exactly that channel's distinguishing power, which is why partial measures such as a
persistent-name transport that pads sizes (or, symmetrically, fresh identifiers without
shaping) cannot reach chance.

\begin{corollary}[The residual is the structural channel]\label{cor:residual}
With unlinkability and metadata minimization but no cover, the only trace-dependent channel
is $N$, so for task class $C$ the recovery adversary's advantage is bounded by the leakage
of that one channel, $\mathrm{I}(C;N)$.
\end{corollary}
\noindent This is the channel the ablation isolates (\S\ref{sec:eval}): the full set,
realized as constant-rate cover, drives recovery to $0.167=1/K$ in \S\ref{subsec:frontier},
the $\mathsf{Adv}=0$ point of Observation~\ref{thm:channels}; dropping cover leaves recovery at
the residual $\mathrm{I}(C;N)$ predicts, the reason a defense short of cover cannot reach
chance. Discovery privacy (Def.~\ref{def:discovery}) is the same game played against the
registry $\mathcal{G}$ rather than $\mathcal{N}$: it holds when the bootstrap view is
independent of the requested capability, the selected peer, and the resulting edge, so that
no $w_0,w_1$ are separable at the discovery layer.

These statements are deliberately elementary; their proofs are immediate from the
definitions. Their role is to make precise \emph{which} channel each property closes and why
the set is necessary, turning ``only as a set'' from an intuition into a coverage argument.
The contribution of this paper is the threat reframing of \S\ref{sec:different} and its
measurement (\S\ref{sec:eval}), and the formalism is a tool for stating that precisely, not a
claim to theoretical depth.

\section{Transport Design Space}\label{sec:transports}
No transport was designed for agent interoperability; each was built for human or
general messaging, so applying it inherits both its protections and its
limitations. Table~\ref{tab:transports} rates candidate transports against the
properties of \S\ref{sec:properties}, alongside the HTTP(S) and SLIM bindings as
baselines. Ratings are qualitative (\emph{strong} / \emph{partial} /
\emph{weak}); the point is the shape of the trade-off, not a score.

\begin{table*}[t]
\centering\footnotesize
\resizebox{\textwidth}{!}{%
\begin{tabular}{@{}lcccccl@{}}
\toprule
Transport & Unlink. & No central obs. & Deniability & Metadata min. & Standardization & Cost \\
\midrule
HTTP(S) / current bindings & weak & weak & weak & weak & IETF, mature & low latency \\
SLIM / SLIMRPC & weak & partial & weak & weak & IETF draft & low latency \\
SimpleX / SMP & strong & strong & strong & partial & non-standard & async, modest \\
Tor onion services & weak & strong & partial & partial & de facto & moderate latency \\
Mixnet (e.g.\ Nym) & strong & strong & strong & strong & non-standard & high latency \\
Oblivious HTTP / relaying & strong & partial & weak & weak & IETF RFC & low latency \\
\bottomrule
\end{tabular}}
\caption{Candidate transports rated against the four wire-transport properties of
\S\ref{sec:properties}; discovery privacy is a bootstrap-layer concern, addressed in
\S\ref{sec:casestudy}. Ratings are qualitative; the point is the trade-off, not a
score. The standardization column marks the axis a binding proposal must also weigh:
SimpleX/SMP rates strongest on the privacy properties yet is non-standard and
single-implementation, while Oblivious HTTP is a deployed IETF standard that rates weak
on metadata minimization, the gap the binding's shaping profile fills.}
\label{tab:transports}
\end{table*}

\paragraph{HTTP(S) bindings (incl.\ SLIM).} Agents are addressed by persistent URL
or structured name, so unlinkability is weak and a network observer or intermediary
obtains the edge directly. SLIM removes the central content-reading broker (an
improvement over a single shared intermediary) but still routes by persistent
name and provides no identifier freshness, mixing, or deniability.

\paragraph{SimpleX / SMP.} Identity-less by construction: connections are
bootstrapped out of band and carried over unidirectional queues with per-queue
identifiers and no global account, giving strong unlinkability and, with separate
and rotating relays, no single observer of the graph; deniability is a design goal.
Its weak point is metadata minimization (a relay still sees the timing and volume
of the queues it hosts), so traffic-analysis defenses are only partial. The model
is asynchronous with modest throughput.

\paragraph{Tor onion services.} Strong at hiding network location and distributing
trust across relays, but a published onion address is a \emph{persistent}
identifier, so unlinkability is weak when agents reuse addresses, and a global
passive adversary can mount traffic correlation, now practical with deep
learning~\citep{deepcorr}. Maturity is high, latency moderate.

\paragraph{Mixnets (e.g.\ Nym).} Purpose-built for metadata protection: per-packet
unlinkable formats, distributed mixing, and cover traffic earn strong ratings on
unlinkability, no-central-observer, and metadata minimization. The cost is high
latency and lower maturity, precisely the trade-off a deployment must weigh.

\paragraph{Oblivious {HTTP} and relaying.} Application-layer designs such as
Oblivious HTTP~\citep{ohttp} and the MASQUE proxying family decouple a request
from the client's identity by interposing two non-colluding parties: a relay that
sees the client but not the content, and a gateway that sees the content but not
the client. Unlinkability of a request to its origin is thus strong, but only if
the two parties do not collude, so no-central-observer is partial. Crucially,
neither party pads, paces, or mixes: the gateway and any network observer still
see message sizes, counts, ordering, and timing, so metadata minimization is weak
and the traffic-analysis channel stays open. The design also targets independent,
low-frequency request/response exchanges and explicitly cautions against linkable,
high-frequency patterns~\citep{ohttp}, which is exactly the correlated multi-step
shape of an agent workflow, and base OHTTP does not carry streaming delivery (its
chunked extension~\citep{chunkedohttp} adds it). It removes
the linkage edge, not the workflow signature.

\paragraph{Takeaway.} No transport provides all four wire-transport properties
cheaply; they trace a privacy/latency frontier. An identity-less messaging fabric
rates strongest on the privacy properties yet sits furthest from current deployment
and review processes; a mixnet is stronger on traffic analysis at a latency cost;
Tor is the most mature but weakest on unlinkability; oblivious relaying unlinks a
request from its origin but does not shape traffic, so the workflow signature
survives. The properties of \S\ref{sec:properties}, not any single transport, are
the portable target.

A caveat of vantage sharpens this. Table~\ref{tab:transports} rates each transport
against an observer \emph{in the network}; a mixnet's strong ratings are earned there.
But any party that handles the \emph{application} messages (a tunnel exit, the
destination agent, a compromised orchestrator, or a content-blind messaging
fabric) still sees their sizes, counts, ordering, and capability labels, none of which
the underlying transport conceals. Metadata minimization must therefore be realized as a
property of the \emph{binding} (\S\ref{sec:casestudy}), not delegated to a transport: a
mixnet hardens the network path, at a latency cost, without on its own closing the
application-layer channel.

\section{Case Study: A Metadata-Protecting Binding for A2A}\label{sec:casestudy}
A2A is a useful case study because it already admits transports beyond its core set
through \emph{custom protocol bindings}, and because its operations are already
asynchronous: an operation returns immediately and updates arrive by polling,
streaming, or push. A metadata-protecting binding is therefore expressible in
principle. We sketch one built by \emph{composition} from standards-track
primitives: an unlinkable carrier (Oblivious HTTP for unary calls~\citep{ohttp} and
its chunked extension~\citep{chunkedohttp} for streamed updates, both over ordinary
HTTP) beneath a
metadata-minimizing shaping layer that realizes the wire properties of
\S\ref{sec:properties} (evaluated in \S\ref{sec:eval}), with capability-scoped
authorization in place of identity-based auth. We prefer composable standards to
transports that rate higher on Table~\ref{tab:transports}, because a binding must be
deployable and reviewable within the existing A2A and IETF processes: the privacy
properties are necessary for a binding, not sufficient for its adoption. The
instructive result is what one meets in trying: even this HTTP-native binding
surfaces three \emph{implicit identity assumptions} the specification never states
because, over plain HTTP, they always hold.

\paragraph{Assumption 1: push notifications assume an HTTP-reachable client.}
A2A push delivers task updates to a client-provided webhook URL~\citep[\S3.1.7]{a2a},
presuming the client has a stable, reachable address; a client unlinkable by
construction has none to offer. This is \emph{surmountable}: server-initiated
delivery re-maps onto client-initiated retrieval, the client polls for updates or
supplies a reply queue it fetches obliviously, or, where low-latency push is required,
a client-established MASQUE-style tunnel~\citep{masque}. Because
A2A already treats push as one of several interchangeable update mechanisms, the
semantics are preserved; only the carrier changes.

\paragraph{Assumption 2: authentication is identity-based.}
A2A authentication is declared in the Agent Card and is identity-bearing (mutual
TLS~\citep[\S4.5.6]{a2a}, OAuth/OIDC, keys tied to a principal). An unlinkable client cannot
present a stable principal without defeating the property the carrier provides. This
is the \emph{genuine mismatch}: schemes that require a verifiable persistent identity
(a client certificate, an OIDC subject) do not translate. What does translate is a
different trust basis: capability-scoped, unlinkable authorization tokens in the
style of Privacy Pass~\citep{privacypass}, optionally bound to the carrier's
handshake. This establishes \emph{what} a peer is entitled to without fixing
\emph{who} it persistently is.

\paragraph{Assumption 3: discovery couples capability to a persistent address.}
A2A discovery resolves an Agent Card at a well-known URL~\citep{a2a} and selects a
binding from its declared endpoint. An oblivious carrier keeps endpoints addressable,
so discovery is the mildest of the three: the card and its endpoint still resolve.
What does not come for free is \emph{unlinkable} discovery: fetching the card and
reaching the endpoint over the oblivious carrier removes the discovery edge from a
network observer, while the capability \emph{content} of the card is unaffected.
Addressing is preserved; only its \emph{observability} gives way.

\paragraph{What the case study shows.} The binding is realizable by composition, but
two of the three assumptions demand real work: push must be re-mapped onto a
client-established tunnel, and identity-based authentication must give way to
capability-scoped, unlinkable authorization; discovery, by contrast, survives an
HTTP-native oblivious carrier almost unchanged. None of the three is stated in the
specification, because over an address-based, identity-bearing transport they hold
for free. Naming them is useful independently of which carrier a binding chooses:
they delimit exactly where interoperability and persistent identity are entangled.
They are also where an adversary would act: discovery and push are the early,
action-coupled steps whose visibility enables the preemption of
\S\ref{sec:different}.

\section{Empirical Evaluation}\label{sec:eval}
Two claims from the earlier sections invite a test. The threat model holds that graph
metadata leaks \emph{pending workflow intent} (\S\ref{sec:different}); the property
framework holds that a defined \emph{set} of properties removes it
(\S\ref{sec:properties}). We evaluate both on real agent traffic, a labeled corpus and a live binding, with a
generative model as a controlled complement, and then ask the third, decision-theoretic
question the integrity framing demands: what the recovered signal is \emph{worth} to
an adversary that acts on it (\S\ref{subsec:actuation}). The findings are that intent is
recoverable from passive metadata, that it is recoverable \emph{early}, and that the
properties reduce recovery sharply toward chance, and only in combination.\footnote{Code and
data: \url{https://github.com/dangoldbj/agent-metadata-privacy}.}

\subsection{Setup}
Our primary evidence is a corpus of real multi-agent A2A traffic,
A2A-MetaTrace, which we build and release (\S\ref{subsec:corpus}). It composes the
official reference \emph{sample} agents, one per capability, each run unmodified as its
own SDK server and backed by real OpenAI and Gemini model calls, into nine workflow
classes, each realized by three distinct capability \emph{compositions}. We record only
$\mathrm{obs}(m)$, and the agents and protocol path are officially authored; only the
composition and the labels are ours, so recovery cannot live in a distribution we
designed. Every headline number below is this measured corpus; we also measure traffic
over a live binding (\S\ref{subsec:binding}) and over MCP (\S\ref{subsec:mcp}).

Two questions a fixed corpus cannot answer call for a controlled complement. Whether the
effect is an artifact of a particular class count, capability overlap, or timing regime
is not testable without varying them; and the decision-theoretic analytics of
\S\ref{subsec:actuation} (Propositions~\ref{prop:kappa}--\ref{prop:alpha}) need balanced
classes and a tunable value model. For these we use a generative model of the A2A task
lifecycle, discovery, delegation, streamed updates, completion, in which each
\emph{task class} is a stochastic process over capability-typed stages drawn from a
shared vocabulary with tunable overlap (so the \emph{set} of capabilities alone does not
identify the class), several agents serve each capability and some are multi-skill (so a
transport-visible identifier is \emph{not} a relabeled capability), and timing and size
profiles attach to capabilities rather than classes (so all class signal flows through
\emph{which} capabilities are invoked and \emph{in what order}). We anchor it to a real
SDK capture and treat it throughout as a controlled instrument, not as a source of
headline magnitudes: where the model and the corpus disagree, the corpus governs.

An \emph{adversary view} projects a trace onto what a single vantage point sees
(Table~\ref{tab:adv}). The \emph{registry} view sees the semantic capability labels
named in discovery queries. The \emph{network} view sees only opaque endpoint
identifiers, timing, volume, and direction, and no semantic labels. A classifier then
predicts the latent task class. Following the website-fingerprinting
tradition~\citep{wfp}, we treat its accuracy above chance as a conservative indicator
of leakage: an unoptimized decoder can only understate it, so we deliberately leave it
untuned. Chance is $1/K$ for $K$ balanced classes, and we
report cross-validated accuracy with a bootstrap $95\%$ confidence interval.

\subsection{Leakage is recoverable from real agent traffic, and prospective}\label{subsec:leakage}
\begin{figure}[t]
\centering
\includegraphics[width=\columnwidth]{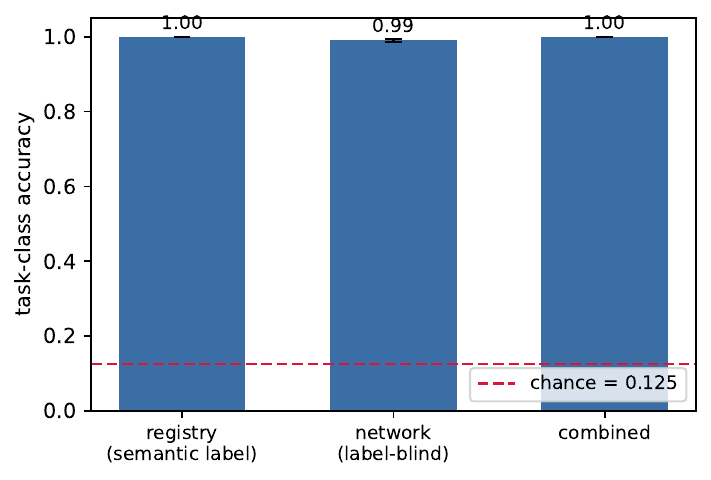}
\caption{Task class recovered from communication-graph metadata, by adversary view, in the
controlled model ($K=8$, chance $0.125$; error bars are bootstrap $95\%$ CIs). Even the
label-blind \emph{network} view, seeing only $\mathrm{obs}(m)$, recovers the class far above
chance; the real-corpus counterpart is $0.67$ (\S\ref{subsec:corpus}).}
\label{fig:leakage}
\end{figure}

On the A2A-MetaTrace corpus (nine classes, chance $0.11$) the \emph{label-blind} network
observer, seeing only $\mathrm{obs}(m)$ and no semantic labels, recovers the task class at
$0.67$, a $6\times$ lift (\S\ref{subsec:corpus} gives the corpus and its
provenance-distinctness). An observer of pure transport metadata is thus far better
informed than chance: capability-correlated timing, volume, and message sequence
reconstruct the capability footprint indirectly, even on this corpus where the endpoints
themselves carry little (\S\ref{subsec:corpus}). This is the semanticity of
\S\ref{sec:different} on real traffic. In the controlled model, where classes are balanced
and timing noiseless, the same label-blind view reaches $0.99$ against the registry view's
tautological $1.00$ (Fig.~\ref{fig:leakage}); the distance from there to the corpus is the
price of real, finite, noisy traffic, and it is the corpus we report.

\begin{figure}[t]
\centering
\includegraphics[width=\columnwidth]{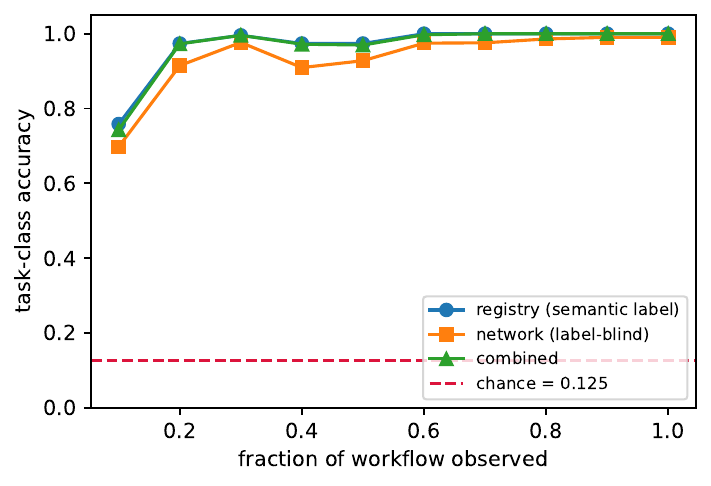}
\caption{Prospectivity in the controlled model: accuracy as a function of the fraction of
the workflow observed. From only its opening, the network view predicts the pending task
class well above chance; on the real corpus the opening yields $0.37$, rising to the full
$0.67$ by the four-tenths mark (\S\ref{subsec:corpus}).}
\label{fig:prospectivity}
\end{figure}

Recovery is also \emph{prospective}. On the corpus the opening predicts the pending class
well before completion: the first three-tenths of a workflow already yield $0.37$ (about
$3.4\times$ chance), and recovery reaches its full $0.67$ by the four-tenths mark. The
controlled model, with its cleaner timing, predicts earlier still, $0.70$ from the first
tenth and climbing toward certainty (Fig.~\ref{fig:prospectivity}). Either way the opening
foretells the trajectory, the prospectivity of \S\ref{sec:different} made concrete and the
precondition for the actuation leverage of \S\ref{subsec:actuation}.

\subsection{The properties neutralize the leak, but only as a set}
We realize the properties of \S\ref{sec:properties} as transforms on the corpus traffic
and re-measure (the wire-faithful observation model of \S\ref{subsec:corpus}, chance
$0.11$). Each wire property alone leaves the workflow recoverable. Fresh identifiers
(\emph{unlinkability}) barely move recovery ($6.0\times\!\to\!6.0\times$): this corpus
carries little class signal in its endpoints, its agents sitting behind one orchestrator,
so closing that channel removes little. \emph{Metadata minimization} (constant cells on a
batched cadence) cuts recovery to $4.2\times$ by closing the timing-and-volume channel that
does most of the work here. Only the two composed drive recovery to chance ($0.5\times$, at
the floor), and the integrity metric falls with them: the capture ratio of
\S\ref{subsec:actuation} drops from $\kappa=0.63$ unprotected to $\approx0$ under the
combined properties. The registry view is a separate channel that neither wire property
touches and that falls only to \emph{discovery privacy} (to exact chance). The threat thus
yields only to the full \emph{set}, each property matched to a channel; partial measures do
not suffice, a content-protecting binding that keeps persistent names being one such case
(\S\ref{sec:problem}).

\begin{figure}[t]
\centering
\includegraphics[width=\columnwidth]{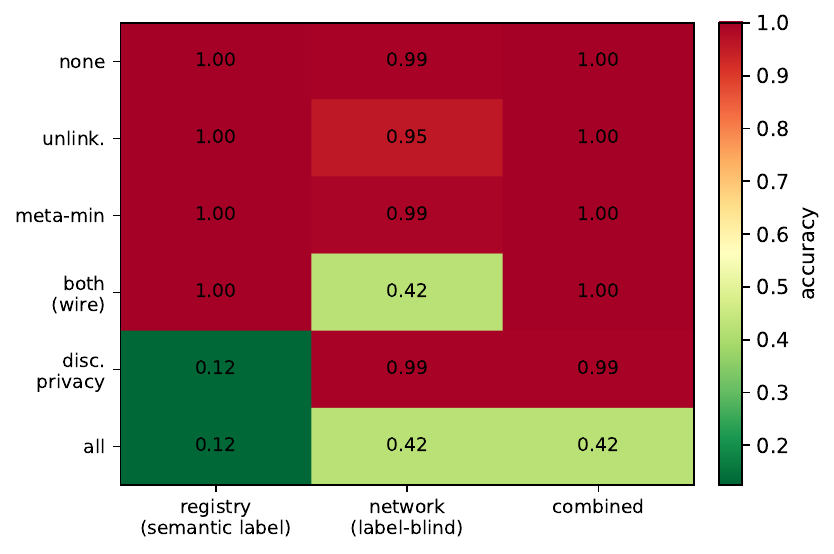}
\caption{Accuracy under each property (rows) for each adversary view (columns) in the
controlled model; red is leaking, green is protected. The network observer falls only when
unlinkability and metadata minimization are combined (``both''); the registry observer falls
only to discovery privacy. No single property suffices. The real-corpus ladder is in
\S\ref{subsec:corpus}.}
\label{fig:protection}
\end{figure}

How much survives the wire properties depends on the observer's granularity, and cover
traffic is what closes the gap wherever a residual remains. The controlled model, with its
sharper per-message structure, shows the same ``only as a set'' pattern but leaves a
residual $0.42$ after both properties (Fig.~\ref{fig:protection}), a \emph{structural}
channel of message counts and sequence shape that padding does not target; the finer-grained
live binding exposes it too, where metadata minimization with fresh identifiers leaves $0.52$
and only constant-rate cover traffic reaches chance (\S\ref{subsec:binding}). The coarse
wire-faithful corpus aggregates each stage's stream into a single observation, so that count
channel is already faint there and the two properties suffice. An oblivious carrier such as
Oblivious HTTP~\citep{ohttp} supplies unlinkability but no shaping, so on it the timing and
volume fingerprint persists, which we confirm on a \emph{live} OHTTP binding
(\S\ref{subsec:binding}).

Why no single property suffices is clearest when the channels are isolated, which we do
on the measured binding capture (\S\ref{subsec:binding}, $K=6$, chance $0.167$). Giving the
classifier \emph{one} channel at a time, each independently recovers the task class far
above chance: endpoint identifiers alone reach $1.00$, inter-message timing $0.84$,
message volume $0.68$, and the bare sequence and counts $0.45$. The
channels are thus \emph{redundant}: removing any one from the full feature set costs
almost nothing ($\le 0.06$), because the others still carry the class. This is the
mechanism behind ``only as a set'', each property closes one channel
(unlinkability the identifiers, metadata minimization the timing and volume, discovery
privacy the registry labels, cover traffic the residual counts), and a defense that
leaves any channel open leaves the redundant signal intact.

\subsection{Actuation: the value of acting on the leak}\label{subsec:actuation}
Leakage and prospectivity are properties of recoverability: how much the metadata
tells an observer about the task. Whether that knowledge bears on workflow
\emph{integrity} is a separate, decision-theoretic question, since an observer that
cannot act on what it learns poses no integrity threat. We therefore model an
adversary that must \emph{act} under a budget and measure what the metadata is worth
to it.

\begin{definition}[Actuation game and value of metadata]\label{def:vom}
Among $N$ concurrent workflows, each $w$ carries an adversary value $v(w)\ge 0$. By a
decision deadline $f$, having observed only the leading fraction $f$ of every
workflow, a metadata-only adversary commits a budget of $B$ interventions, choosing a
set $S$ with $|S|=B$ to maximize $J(S)=\sum_{w\in S} v(w)$, and ranks workflows using
the label-blind network view of the observed prefix alone. With $J_{\mathrm{inf}}$,
$J_{\mathrm{blind}}$, and $J_{\mathrm{orc}}$ the objective under the
metadata-informed, uniformly random, and true-value selections, the \emph{value of
metadata} is $\mathrm{VoM}(B,f)=J_{\mathrm{inf}}-J_{\mathrm{blind}}$ and the
\emph{capture ratio} is
$\kappa=(J_{\mathrm{inf}}-J_{\mathrm{blind}})/(J_{\mathrm{orc}}-J_{\mathrm{blind}})$,
normalized so that $0$ is the blind baseline and $1$ the oracle; a ranking worse than
random can fall below $0$, and in our experiments $\kappa\in[0,1]$. It is the share of
the attainable advantage the adversary realizes from metadata alone.
\end{definition}

We instantiate the game minimally, adding no new dynamics to the workflows. One task class is
the adversary's target, $v(w)=1$ if $w$ is of that class and $0$ otherwise; the budget
equals one class's mass; the ranking is the network observer's out-of-fold posterior
on the target, from exactly the prefix features used in the preceding subsections. Then
$J_{\mathrm{inf}}$ is the count of true target workflows among the top-$B$ by that
posterior, and $J_{\mathrm{blind}}$ and $J_{\mathrm{orc}}$ are closed-form; we average
over the choice of target. What $\mathrm{VoM}$ and $\kappa$ capture is \emph{selection}
leverage: the advantage of picking the right workflows to act on, from the opening
alone. That is the precondition for changing outcomes, not proof that they change;
the latter would mean acting against a live binding (\S\ref{sec:discussion}).

This instantiation makes $\kappa$ analytically transparent, and the form explains the
defense behavior we will observe. Write $p$ for the target class's prevalence (its mass as a
fraction of the population), let the budget be that mass $B=pN$, and let $\mathrm{Prec}@B$ be
the \emph{precision} of the adversary's top-$B$ ranked set, the fraction of those $B$
workflows that are truly targets.

\begin{proposition}[Capture ratio is governed by top-$B$ precision]\label{prop:kappa}
For the unit-value game with the budget equal to the target mass, the capture ratio depends on
the adversary's ranking \emph{only} through its top-$B$ precision:
\[
  \kappa \;=\; \frac{\mathrm{Prec}@B - p}{1 - p}.
\]
In particular $\kappa$ does not depend on the adversary's overall classification accuracy; a
defense that drives the precision of the top-ranked workflows toward the base rate
($\mathrm{Prec}@B\!\to\!p$) drives $\kappa\!\to\!0$ even if bulk accuracy remains well above
chance, and a ranking worse than random ($\mathrm{Prec}@B<p$) gives $\kappa<0$.
\end{proposition}
\noindent The argument is immediate from the definitions: the informed objective is
$J_{\mathrm{inf}}=B\cdot\mathrm{Prec}@B$, the blind objective is $J_{\mathrm{blind}}=B\,p$, and
since the budget equals the target mass the oracle takes every target, $J_{\mathrm{orc}}=B$;
substituting into Definition~\ref{def:vom} gives the stated form. The consequence is the
key structural fact of the actuation axis: \emph{selection leverage tracks precision at the
budget, not accuracy over the population}. It is why, below, the combined wire properties
collapse $\kappa$ to the blind baseline even though the label-blind observer still recovers
task class at $0.42$: padding and fresh identifiers flatten the top of the adversary's
posterior, where selection under a tight budget lives, while leaving enough bulk signal to
label the easy majority. The integrity objective ($\kappa$) and the privacy objective
(recovery) are therefore not the same target, and a defense tuned to one need not move the
other.

\begin{figure}[t]
\centering
\includegraphics[width=\columnwidth]{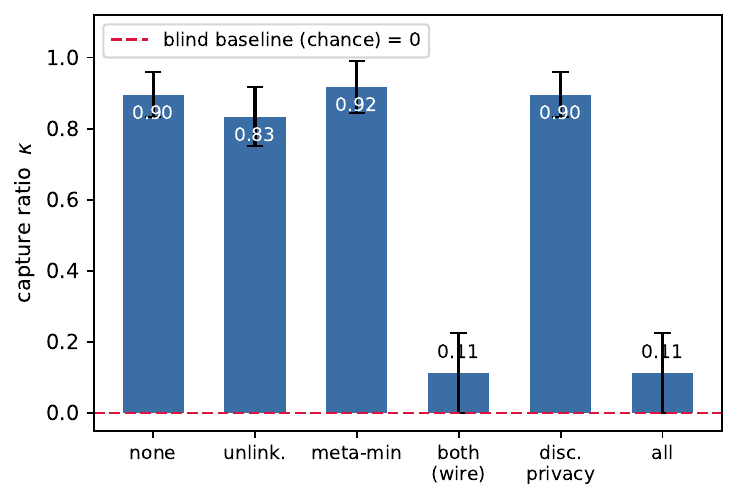}
\caption{Actuation. Capture ratio $\kappa$ by privacy property, at an early decision
deadline ($f=0.2$) and a budget equal to one task class's mass, averaged over targets
(error bars span $\pm1.96$ standard errors across target classes; chance, the blind
baseline, is $0$). The
integrity analogue of Fig.~\ref{fig:protection}: only the combined wire properties
(``both'') collapse the leverage, and discovery privacy, which the label-blind
observer ignores, does not.}
\label{fig:actuation}
\end{figure}

\begin{figure*}[t]
\centering
\includegraphics[width=\textwidth]{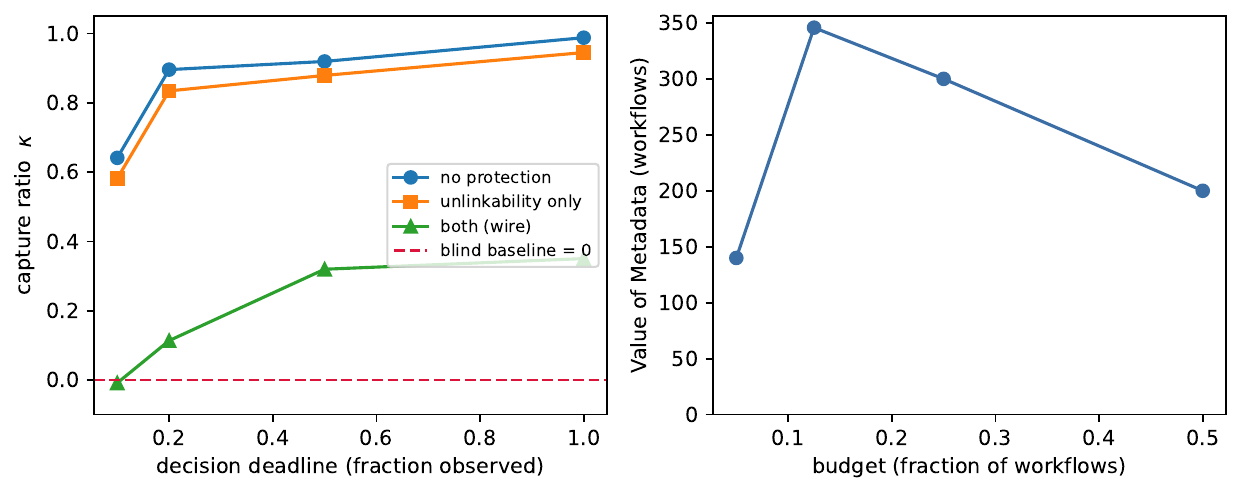}
\caption{Actuation is the product of inference and budget, vanishing on either edge.
Left: capture ratio against the decision deadline; leverage tracks prospectivity, is
substantial even from a short prefix, and the combined wire properties hold it at the
blind baseline. Right: value of metadata against budget (no protection, full
workflow); it vanishes without a budget to spend and peaks where the budget is scarce
relative to the target set, the signature of a value-of-information quantity.}
\label{fig:actsep}
\end{figure*}

The value of metadata is substantial and, like the inference beneath it, prospective.
On the corpus, an adversary deciding from only the opening fifth of each workflow captures
$\kappa=0.41$ of the oracle's advantage over the blind baseline, rising to $0.63$ once the
full workflow is seen, with the budget set to one class's mass. The controlled model, where
the opening is cleaner, captures more and earlier ($\kappa\approx0.90$ from the opening fifth,
Fig.~\ref{fig:actuation}, climbing toward unity, Fig.~\ref{fig:actsep}, left); the corpus is
the conservative figure. Either way, under a budget, knowing the task early is most of the
way to acting on it.

Actuation is not a restatement of leakage. The value of metadata is the product of two
independent factors (an early-decidable signal and a budget to spend), and
collapses if either is absent: with no budget there is nothing to actuate, so
$\mathrm{VoM}\to0$ as the budget shrinks (Fig.~\ref{fig:actsep}, right), and with no
early signal the ranking is uninformative and $\kappa\to0$. The axis is genuinely
separate from recoverability, and website-fingerprinting, which bounds the signal
factor alone, does not speak to it.

The defense carries over on the corpus, and again only as a set. Each wire property alone
leaves the leverage substantial ($\kappa=0.61$ under unlinkability, $0.50$ under metadata
minimization, at the full deadline); the two together drive $\kappa$ to $\approx0$, the blind
baseline, and cover with fresh identifiers holds it there. Selection leverage is downstream of
inference: once the combined properties take recovery to chance, there is nothing left to
target. The controlled model lets us see the sharper form of this that
Proposition~\ref{prop:kappa} predicts, where the two objectives \emph{come apart}: there the
combined properties leave the label-blind observer still recovering task class at $0.42$
(Fig.~\ref{fig:protection}), yet drive $\kappa$ to $0.12$ (Fig.~\ref{fig:actuation}), because
padding and fresh identifiers flatten the top of the posterior, where selection under a tight
budget lives, while leaving enough bulk signal to label the easy majority. This come-apart is
not an artifact of the model's balanced classes and noiseless timing: it recurs on the live
binding, where on real \texttt{a2a-sdk} wire timing the combined properties hold task-class
recovery at $0.375$ (above the $0.167$ chance) yet drive $\kappa$ to $0.02$, the blind baseline
(\S\ref{subsec:binding}). Wherever the structural count channel survives the wire properties it
supports bulk labeling but not the top-ranked precision that budgeted selection needs. That
measured $\kappa\!\approx\!0.12$ in the model corresponds to a top-$B$ precision close to the
base rate, the analytic prediction made concrete. Discovery privacy, which does not touch the
network view, leaves $\kappa$ unchanged, as expected. The integrity defense
thus follows from the privacy defense, since the properties that suppress what an
observer can recover suppress the leverage it gains.

\subsection{Actuation, demonstrated: a live front-running suite}\label{subsec:demo}
The capture ratio quantifies the leverage; we now show it is realizable as a
running exploit, and removed by the same property. In a closed testbed a victim
repeatedly runs a two-step A2A workflow over the real SDK: an \emph{opening} against a
worker agent whose streamed footprint depends on the workflow's latent type, then an
\emph{action} against an agent that grants each scarce slot to whoever asks first. A
passive observer sees only the opening's $\mathrm{obs}(m)$, never a payload; from a short
prefix it predicts the type and races a competing action \emph{before the victim reaches
that stage}, a real wall-clock front-run. We run three scenarios spanning two kinds of
leverage. In \emph{resource race} (selection) the adversary predicts \emph{whether} a
workflow is high-value and, under a budget, races a single slot; in \emph{discovery
poisoning} and \emph{bid preemption} (targeting) it predicts \emph{which} of $K$
capabilities or items the workflow will act on and races that target.

The results are stark and uniform (Table~\ref{tab:demo}, $100$ rounds per condition).
Over HTTPS the observer front-runs essentially every high-value workflow (capture
$1.00$, precision $1.00$); under the metadata-minimization shim, with the opening padded
to constant cells, capture falls to the blind baseline in all three ($0.53$ vs.\ $0.50$;
$0.26$ vs.\ $0.25$; $0.18$ vs.\ $0.17$), and the value of metadata collapses from
$25$--$83$ captured workflows to within noise of zero. The shim does not stop the race;
it destroys the prediction that aims it, whether the adversary was selecting which
workflows to act on or targeting which action to preempt. This is the integrity threat as
a demonstration rather than a model, and the defense of \S\ref{sec:properties} measured
against it. The testbed is closed (our agents, a constructed scarce resource), so this
is a running exploit, not a field measurement; but it is a race in real time, not
arithmetic over a population.

The structure of the cost is what matters: in a bid-preemption setting, an adversary that
front-runs the bid it predicts from a workflow's opening wins the contested item on
essentially every high-value round ($1.00$) against $0.17$ for a blind competitor
(Table~\ref{tab:demo}), and the metadata-minimization property removes that edge. The swung
value scales as the metadata advantage times the value at stake times the workflow rate; that
product, rather than any single dollar figure, is what makes a metadata leak an
\emph{integrity} cost rather than a privacy one, and it grows as agent commerce moves to this
substrate.

\begin{table}[t]
\centering
\resizebox{\columnwidth}{!}{\small
\begin{tabular}{llccc}
\toprule
scenario & leverage & HTTPS & shim & blind \\
\midrule
resource race      & selection & $1.00$ & $0.53$ & $0.50$ \\
discovery poison   & targeting & $1.00$ & $0.26$ & $0.25$ \\
bid preemption     & targeting & $1.00$ & $0.18$ & $0.17$ \\
\bottomrule
\end{tabular}}
\caption{Live front-run: fraction of high-value workflows the metadata-only adversary
captures, by transport ($100$ rounds each). Under HTTPS it front-runs nearly all;
under the shim it falls to the blind baseline in every scenario, across both selection
and targeting leverage.}
\label{tab:demo}
\end{table}

\subsection{General actuation: value distributions and competition}\label{subsec:general}
Proposition~\ref{prop:kappa} is exact but special: a unit value on one class and a
budget equal to that class's mass, where $\kappa$ reduces to top-$B$ precision. That
reduction invites the reading that the integrity axis is precision under another name.
It is not, and the general game (Definition~\ref{def:vom}, which already admits arbitrary
$v(w)\ge0$ and any budget) shows why. We generalize along the three axes the special case
leaves open: arbitrary value distributions, multiple budgets, and competing adversaries.
These analytics need a tunable value model, so we work in the controlled model; the same
out-of-fold posteriors as before drive every result, and only the value model and the field
of adversaries change. We report the early-deadline, headline-budget cell and average over
random value assignments; the full sweep over budgets is monotone and unsurprising and we
omit it.

The first question is what a metadata-only adversary can even rank by when value is not a
class indicator. Its value-maximizing score is the posterior-expected value
$\mathbb{E}[v\mid \mathrm{obs}]$, and this is governed by how much of the value the leaked
class signal explains.

\begin{proposition}[The value of metadata scales with the class-determined share]\label{prop:alpha}
Let value decompose as $v(w)=\alpha\,\mu_{c(w)}+(1-\alpha)\,\varepsilon_w$, where $c(w)$ is
the task class, $\mu_c=\mathbb{E}[v\mid c]$, and the idiosyncratic part $\varepsilon_w$ is
independent of $\mathrm{obs}(w)$. The metadata adversary's value-maximizing score is
$\mathbb{E}[v\mid \mathrm{obs}]=\alpha\sum_c \Pr(c\mid \mathrm{obs})\,\mu_c+(1-\alpha)\bar\varepsilon$,
whose ordering is that of the class-value posterior \emph{for every $\alpha>0$}, so the
adversary selects the same set $S$ at every $\alpha$. Its value of metadata is then linear in
that share:
\[
  \mathbb{E}\,[\mathrm{VoM}_{\mathrm{inf}}(\alpha)] \;=\; \alpha\cdot\mathbb{E}\,[\mathrm{VoM}_{\mathrm{inf}}(1)],
\]
and since the oracle, which sees $\varepsilon$, retains advantage as $\alpha\to0$ while the
adversary's vanishes, $\kappa(\alpha)\to0$. Metadata buys leverage only to the extent value
is class-determined; a workflow whose stakes are idiosyncratic to its type is unrankable from
the graph however well the type is recovered.
\end{proposition}
\noindent The score is linearity of expectation applied to the decomposition; its ordering is
$\alpha$-invariant because the $\alpha$ scaling and the additive constant do not reorder, so
$S$ is fixed. On that fixed set the class term contributes $\alpha(\sum_S\mu-B\bar\mu)=
\alpha\,\mathrm{VoM}_{\mathrm{inf}}(1)$ and the idiosyncratic term is mean-zero, since $S$ is
chosen by class and $\varepsilon\perp\mathrm{class}$. Proposition~\ref{prop:kappa} is the
corollary $\alpha=1$, $v=\mathbf{1}[c=t]$, budget the target mass, where the closed form returns.

\begin{figure*}[t]
\centering
\includegraphics[width=\textwidth]{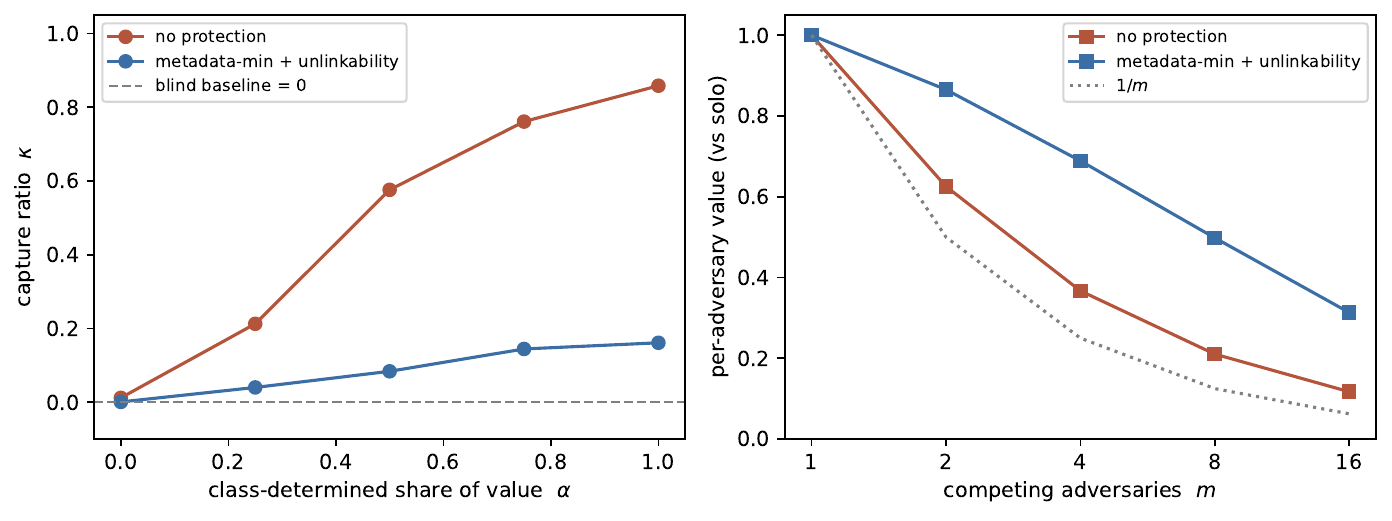}
\caption{General actuation. Left: capture ratio against the class-determined share of
value $\alpha$ (heavy-tailed values, early deadline, one-class-mass budget); leverage rises
with how much of the value the leaked class signal explains (Proposition~\ref{prop:alpha})
and the combined wire properties hold it at the blind baseline throughout. Right: per
adversary value against the number of competitors sharing the signal, normalized to the
solo take; actuation leverage is rivalrous and dilutes roughly as $1/m$, unlike a
non-rivalrous privacy leak.}
\label{fig:actgen}
\end{figure*}

The empirics bear this out (Fig.~\ref{fig:actgen}, left). With heavy-tailed but
class-determined stakes ($\alpha=1$, per-class values drawn lognormal), the leverage is
near its unit-value level under no protection ($\kappa=0.94$) and the combined wire
properties drive it to $0.09$: $\kappa$ is well above zero for a non-unit value, so it is
not merely a precision count, and it still collapses. Sweeping the determinacy knob with
value only partly tied to class ($\alpha$ from $1$ to $0$) traces the predicted decline:
$\kappa$ falls
$0.86\to0.76\to0.58\to0.21\to0.01$ as value decouples from the leaked signal, while the
defense holds it at or below $0.16$ at every $\alpha$. The integrity threat is real exactly
when high value is metadata-predictable, and the defense neutralizes it across the range.

The second axis is competition. A privacy leak is non-rivalrous: any number of observers can
each recover the task class. Actuation is not, because acting consumes a scarce slot, so the
advantage is competed away. Modeling $m$ adversaries that share the signal and split the
value of any workflow they both select, the per-adversary take dilutes roughly as $1/m$
(Fig.~\ref{fig:actgen}, right): the leak's worth to any one adversary erodes as the field
grows, even as the aggregate damage to the victim population does not. This is the
machine-speed-market reading of the MEV analogy (\S\ref{sec:different}) made quantitative,
and it marks a structural difference between the privacy and integrity objectives rather
than a restatement of the leakage number.

This sharpens the defense-design consequence: a $\kappa$-optimal transport should
flatten the precision of the adversary's top-ranked workflows, where selection under a tight
budget lives, rather than minimize bulk recovery. The two objectives coincide here because
the combined properties do both, but Proposition~\ref{prop:kappa} shows they need not, and a
defense could in principle leave bulk accuracy high while killing the leverage.

\subsection{Onto a real wire: a live binding}\label{subsec:binding}
The corpus fixes the leak on real workflows and the model lets us sweep defenses, but
neither yet carries a defended workflow over a real transport, where latency and shaping
costs are decided. We do that here: holding a workflow population fixed, we replace each
message's $(t,\ell)$ with values \emph{measured on a real \texttt{a2a-sdk} round-trip},
discovery, JSON-RPC \texttt{message/send}, SSE-streamed updates, completion, carried over a
chosen transport. The capture records only $\mathrm{obs}(m)$ and discards bodies, so even a
loopback capture is faithfully a passive TLS observer's view, and the encrypted-record
size tracks plaintext plus a near-constant overhead. This view is invariant to the
content-encryption scheme: TLS, MLS, and SLIM all encrypt the payload and leave the
same $(\text{endpoints},t,\ell,d)$ exposed, so the result holds under A2A's MLS binding
too. We run the same label-blind network classifier ($K=6$, chance $0.167$) over four
transports and measure both leakage and the latency it costs
(Table~\ref{tab:binding}).

\begin{table}[t]
\centering
\resizebox{\columnwidth}{!}{\small
\begin{tabular}{lccc}
\toprule
transport & net.\ leakage & + fresh ids & latency/wf \\
\midrule
HTTPS direct        & $0.99$ & $0.92$ & $0.26$\,s \\
Tor (onion)         & $0.98$ & $0.65$ & $5.89$\,s ($22\times$) \\
Nym (live mixnet)   & $0.97$ & $0.73$ & $8.26$\,s ($32\times$) \\
metadata-min shim   & $1.00$ & $\mathbf{0.32}$ & $1.62$\,s ($6\times$) \\
\bottomrule
\end{tabular}}
\caption{Label-blind task-class recovery and per-workflow latency, measured over a
live A2A binding ($K=6$, chance $0.167$). ``+ fresh ids'' composes unlinkability with
each transport. Naive anonymity costs the most and helps the least: Tor pays $22\times$
latency yet still leaks the class, and a real mixnet, at $32\times$, barely moves
recovery to an app-edge observer. The purpose-built metadata-minimization shim, composed
with fresh identifiers, drives recovery below even the live mixnet, at a
fraction of its latency (and $\sim\!4.8\times$ bandwidth, Table~\ref{tab:frontier}),
though reaching \emph{exact} chance needs the constant-rate cover traffic of
\S\ref{subsec:frontier}.}
\label{tab:binding}
\end{table}

The measured picture matches the analysis: on real \texttt{a2a-sdk} timing and volume the
label-blind observer recovers the task class at $0.99$ at this $K$, so the leak is carried by
wire-level timing and volume, not by any artifact of how workflows are generated, and the
corpus has already shown it on real workflows. The defense story sharpens. General-purpose anonymity networks are the wrong layer: Tor
anonymizes the path but preserves per-message timing, size, and direction, and a mixnet
perturbs timing but barely dents recovery even at a latency no interactive workflow can
bear, while sizes and message counts survive below it. The same holds for an
application-layer tunnel such as MASQUE, the carrier the case-study binding uses for
streamed updates (\S\ref{sec:casestudy}): it relays the stream without reshaping it, so
the shaping must be applied \emph{over} the tunnel, not delegated to it. Defenses at
adjacent layers fall short for the same reason: content- or {DNS}-level obfuscation dampens
but does not remove the metadata signal~\citep{netleak}, so the shaping must live in the
binding. The
metadata-minimization shim, which pads every wire unit to a constant cell on a fixed
cadence and composes with fresh identifiers, drives recovery to $0.32$, about
$1.9\times$ chance and below the live mixnet at a fraction of its latency, though
closing the residual to \emph{exact} chance still requires the constant-rate cover
traffic of \S\ref{subsec:frontier}. Padding sizes
and pacing alone leaves the persistent-identifier channel ($0.996$); only composing fresh
identifiers collapses recovery, the ``only as a set'' lesson on measured traffic.

\paragraph{The standard one would reach for.} The most natural objection is that an
existing IETF standard, Oblivious HTTP~\citep{ohttp}, already addresses this. We
implemented it to find out. OHTTP decouples a request from the client's identity through
a non-colluding relay and gateway, realizing unlinkability (Def.~\ref{def:unlink}) and
discovery privacy (Def.~\ref{def:discovery}) at the application layer, but it shapes no
traffic; and because it carries no server-streamed updates, the binding obtains them by
polling at a fixed cadence. Run as a live binding over the real \texttt{a2a-sdk} (client
to relay to gateway to agent, with real HPKE encapsulation), bare OHTTP drives the
discovery channel to chance ($1.00\!\to\!0.17$) yet still recovers the task class at
$0.65$, roughly four times chance: unlinkability and discovery privacy leave the timing,
volume, and sequence channels open. Composing the metadata-min shim, the configuration the
case study proposes, lowers recovery to $0.52$ but not to chance: the shim equalizes size,
yet the number of messages a workflow emits still tracks its length, and so its class, the
\emph{structural} channel of \S\ref{subsec:frontier} that only constant-rate cover closes.
Adding that cover, padding every workflow to a constant stage count with fresh per-session
identifiers (the binding's full discipline), drives live recovery to $0.30$, at roughly
$13\times$ the baseline bytes and $11\times$ its latency. The residual above chance is a
sub-millisecond timing signal, an artifact of a loopback testbed that co-locates client,
relay, gateway, and agent on one host: ablating the timing features alone collapses recovery
to exact chance ($0.167$), the floor the idealized frontier reaches (\S\ref{subsec:frontier}),
and a signal that real-network jitter, being class-independent, would mask. The fresh
identifiers leave the endpoint channel with no measurable class signal of its own. The
standard a deployer would reach for first thus supplies two of the five properties; closing
the size channel needs the shaping the binding adds, and reaching chance needs cover traffic,
which the live binding realizes. The binding and these measurements are in the artifact.

\subsection{A defense-aware adversary}\label{subsec:adaptive}
The classifier so far is fixed and unoptimized, by design a lower bound on leakage. A
defense claim, though, must hold against an adversary that \emph{adapts} to the defense; the
website-fingerprinting line is a cautionary precedent, where deep-learning attacks reopened
defenses once believed adequate~\citep{df, kfp, varcnn}.
We therefore give the adversary the protected traffic to train on, features that
constant-cell padding cannot erase, message and segment counts, direction-run structure,
the per-stage cell tallies the cadence leaves intact, and an in-fold model and
hyperparameter search (nested cross-validation, so no test row informs selection). This
adversary does not overturn the defense. On the shim with fresh identifiers, recovery
moves only from $0.32$ to $0.36$ against chance $0.167$; on the post-hoc combined wire
properties it holds near $0.37$ (a defense-aware $0.371$ against the fixed $0.375$), about
twice chance, and on unprotected HTTPS it gains nothing. What survives is precisely the \emph{structural} channel,
message counts and sequence shape, that metadata minimization by definition does not
target ($(t,\ell,d)$, not how many cells flow); the same residual appears identically
across Tor, the shim, and matched HTTPS, confirming it is the workflow's plan, not the
transport, that leaks last.

\subsection{The cost of the frontier}\label{subsec:frontier}
Closing that last channel has a price, and cheaper defenses do not pay it. We sweep a
frontier of schemes against the adaptive adversary and record the bandwidth each costs
(Table~\ref{tab:frontier}), drawing the schemes from the website-fingerprinting-defense
literature: size-bucket padding, the zero-delay dummy injection of
FRONT~\citep{front} and the adaptive padding of WTF-PAD~\citep{wtfpad}, and the constant-rate
regime of Tamaraw~\citep{tamaraw} and Surakav~\citep{surakav}. Padding sizes to power-of-two
buckets, or injecting FRONT-style dummy bursts, barely moves recovery; the combined wire
properties drop it to
$0.37$ but leave the count channel; only constant-rate cover traffic with fresh
identifiers, in which every workflow is rewritten to one indistinguishable cell stream,
reaches exact chance, at $16\times$ the baseline bytes. The frontier is monotone and
unforgiving: there is no cheap point that closes the structural channel.

\begin{table}[t]
\centering
\resizebox{\columnwidth}{!}{\small
\begin{tabular}{lcc}
\toprule
scheme & leakage (adaptive) & bandwidth \\
\midrule
none                      & $0.99$ & $1.0\times$ \\
size-bucket padding       & $0.99$ & $1.5\times$ \\
FRONT-style dummies       & $0.98$ & $1.2\times$ \\
wire properties (both)    & $0.37$ & $4.9\times$ \\
constant-rate cover       & $0.23$ & $16.2\times$ \\
\quad + fresh identifiers & $\mathbf{0.17}$ & $16.2\times$ \\
\bottomrule
\end{tabular}}
\caption{Cost--leakage frontier on the measured HTTPS capture, evaluated against the
defense-aware adversary of \S\ref{subsec:adaptive} (chance $0.167$). Cheap padding buys
almost nothing; only constant-rate cover with fresh identifiers reaches chance, and only
at substantial bandwidth.}
\label{tab:frontier}
\end{table}

\subsection{Partial adoption}\label{subsec:partial}
A metadata-protecting binding is a deployment property, and deployment is incremental.
We model an ecosystem in which a fraction $\rho$ of serving parties adopt the wire
properties and the rest keep plain HTTPS, applying the protection only to messages that
touch an adopting party. Recovery falls slowly and late: at $\rho=0.5$ it is still
$0.91$, at $\rho=0.75$ still $0.79$, and only at full adoption does it reach the
protected $0.375$; at $\rho=0.25$ it does not fall at all, since a partially protected
workflow is itself a distinguishable pattern. Metadata privacy here has the character of
a herd property, weak until adoption is near-total, which bears directly on how such a
binding should be introduced into a standard (\S\ref{sec:discussion}).

\subsection{Distinctiveness: is agent metadata more revealing?}\label{subsec:distinct}
The threat model (\S\ref{sec:different}) claims agent metadata is distinctively
\emph{consequential}, and does not claim it is more \emph{revealing} than other
structured machine traffic. We test that revealing claim directly. Using the same pipeline and a matched design (nine
classes, $270$ traces each, the length-free common feature set), we compare the real-agent
corpus (\S\ref{subsec:corpus}) against production microservice call graphs from a public
cluster trace~\citep{alibabatrace}, taking each call graph's root service as its class label,
the analogue of a task class in that the root entry point governs the downstream sequence of
service calls just as a task class governs its capability sequence. The comparison favors
neither side by construction: both run the identical pipeline and feature set, with classes
and trace counts matched. Microservice traffic is recovered \emph{at least} as well as agent
traffic: $0.92$ versus
$0.66$ for agents (chance $0.11$; lifts $8.3\times$ and $6.0\times$), and at a short prefix it
is markedly more predictable ($0.87$ versus $0.37$ at a tenth of the workflow). Agent metadata
is thus, if anything, \emph{less} revealing than a service mesh on the common feature set, not
more. The one structural difference is modest and in the familiar direction, agent recovery
leans less on raw trace length (a length-only baseline reaches $0.31$ for agents but $0.42$
for microservices, and under length-stratified folds agent recovery barely moves,
$0.66\rightarrow0.69$).

This confirms the framing rather than undercutting it. Inference strength
is \emph{not} where the agent setting departs from decades of traffic analysis; a
service mesh fingerprints about as well. What differs is who is positioned to perform
the inference, an outside party across a trust boundary rather than the cluster operator
who already holds the data, and what acting on it yields, leverage over a
machine-speed action rather than a read on an internal call. The contribution rests on
the vantage and actuation axes of \S\ref{sec:different}, which this comparison leaves
untouched, not on a claim of superior recoverability, which it refutes.

\subsection{A provenance corpus, and the limit of recovery}\label{subsec:corpus}
The headline results run on A2A-MetaTrace, the corpus we describe here in full and release.
It is built from the official reference \emph{sample} agents (one per capability, each run
unmodified as its own server over the SDK and backed by real OpenAI and Gemini language-model
calls), composed by an orchestrator into nine workflow classes, each realized by three distinct
capability \emph{compositions}. We record only $\mathrm{obs}(m)$, and the workflow composition
is no longer ours to tune per class; the protocol path and the agents are officially authored,
only the composition and the labels are ours, so recovery cannot live in a distribution we
designed. On this provenance-distinct traffic ($270$ workflows, chance $0.11$) the label-blind
observer recovers the class at $0.67$, a $6\times$ lift, and prospectively: the first
three-tenths of a workflow yield $0.37$ (about $3.4\times$ chance) and recovery reaches its
full $0.67$ by the four-tenths mark (\S\ref{subsec:leakage}).

The recovery is structural, not an artifact of one chatty agent. Because a network observer
sees TLS-record bursts rather than the per-event deltas a streaming model emits, we also score
a wire-faithful view that aggregates each stage's streamed response into a single observation;
recovery is essentially unchanged ($0.68\rightarrow0.67$ as the mean messages per workflow fall
from $376$ to $11$), so it rests on workflow structure (which capabilities, in what order, at
what response volume) rather than on streaming volume. And the same metadata-protecting
properties close it, only as a set: against this wire-faithful adversary fresh identifiers
alone barely move recovery ($6.0\times\!\to\!6.0\times$, the endpoint channel being faint on
this orchestrator-fronted corpus), metadata minimization cuts the lift to $4.2\times$, the two
composed reach chance, and cover traffic with fresh per-workflow identifiers holds it there.
The integrity metric tracks the same ladder: the capture ratio of \S\ref{subsec:actuation}
falls from $\kappa=0.63$ unprotected (budget one class's mass) to $\approx0$ under the combined
properties, so acting on the leak is worth nothing once it is defended.

The corpus also bounds what the recovery is, and the bound is sharp. The signal lives at
the granularity of the \emph{recurring workflow}, not an abstract task intent: of the nine
classes, each is realized by three capability compositions ($27$ in total), and a classifier
asked to name the $27$-way composition does so as well as it names the $9$-way class
($0.68$ versus $0.67$), so class recovery is composition recognition followed by grouping.
The decisive test is holding out whole compositions (leave-one-composition-out, so the
adversary is scored on a workflow shape it never trained on): recovery collapses to chance
($0.18$, against $0.11$), and \emph{seven of the nine classes fall to zero recall}, the
adversary unable to recognize a held-out composition as its class even from the two sibling
compositions of that class it did train on. The label-blind adversary thus recognizes
\emph{specific, recurring} workflows; it does not infer the intent of a novel one. We
therefore state the threat as \emph{recurring-workflow recognition}, not zero-shot intent
inference. This is the realistic model, not a weaker one: the adversary is not a
zero-shot oracle but one that observes a deployment over time and acts on
\emph{recognition} of its recurring workflows, which is exactly the regime where recovery
is strong. It also bounds the claim, against genuinely novel one-off workflows, graph
metadata alone does not reveal the task.

\subsection{Beyond A2A: the same leak over MCP}\label{subsec:mcp}
The threat is stated for the communication graph, not for A2A in particular, so it should
transfer to other interoperability protocols. We check the most prominent, the Model
Context Protocol, over its \emph{real} streamable-HTTP transport with the official SDK: a
FastMCP server exposes several tools with distinct response footprints, a client drives
constructed workflow classes (each a fixed tool sequence), and an ASGI middleware records
only $\mathrm{obs}(m)$ on the wire. This is a transport-generality check rather than a second
agent corpus; the workload is synthetic, so recovery here reflects clean constructed classes
as in the controlled model, not a real-traffic magnitude. The test it makes is structural:
MCP streamable HTTP exposes a \emph{single} server endpoint, so the persistent-identifier
channel is absent and any recovery must rest on volume, timing, and sequence alone. It still
does, over $200$ workflows in five classes the label-blind network view recovers the class at
$0.99$ (chance $0.20$), so even without endpoints the metadata leaks the class. The leak is
thus a property of the metadata any address-based transport exposes, not of A2A, and the
defense is a transport-layer concern common to the interoperability stack.

\subsection{Robustness and scope}
The effect is structural rather than tuned. Across sweeps of the number of classes,
the capability overlap, and the timing noise, the network view stays $4$--$14\times$
above chance, the short-prefix prediction stays above chance, and the two wire
properties always collapse recovery. We therefore claim the \emph{structure} of the
effect, not its precise magnitude; this holds for the value of metadata too. The simulated
sweeps are anchored at both ends by measured traffic: the real-agent corpus
(\S\ref{subsec:corpus}) reproduces the leak ($6\times$ chance), its robustness to a
wire-faithful observation model, and its collapse to chance under the defense, on official-SDK
agent traffic; and the live binding (\S\ref{subsec:binding}) shows the same leakage and defense
structure on the wire. Actuation is measured on the corpus as well as the model: a budgeted
adversary's selection captures $\kappa=0.63$ of the oracle advantage on the real corpus
($0.41$ from the opening), driven to the blind baseline by the combined properties. What
stays in the model is the sharper Proposition-\ref{prop:kappa} regime, in which $\kappa$
collapses while bulk recovery persists; what stays open is an end-to-end exploit against a
deployed binding, which \S\ref{subsec:demo} realizes as a live race in a closed testbed. The
evaluation thus substantiates all axes on real traffic, semanticity, prospectivity, and
actuation, with end-to-end manipulation against a production binding the natural next step.

\section{Discussion}\label{sec:discussion}

\subsection{Reputation and trust without a global graph}
A natural objection, especially in the current agent ecosystem, is that trust and
reputation require persistent identity and an observable history of interactions,
which metadata privacy appears to break. The tension is real but narrower than it
first appears. It conflicts only with \emph{global-observation} reputation: a
registry or ledger that watches all interactions to compute scores. That design is
fundamentally incompatible with unlinkability, and is itself a graph-surveillance
mechanism, i.e.\ the very asset of \S\ref{sec:threat}. It is compatible, however,
with two other models. Under \emph{credential-based} reputation, portable signed
attestations are selectively disclosed: an agent proves ``a verifier attests that I
completed $N$ tasks at quality $q$'' without revealing whom it transacted with, so
trust travels with the agent rather than being reconstructed from observed traffic.
Under \emph{pairwise} reputation, two agents accrue trust over their own repeated
interactions with no global observer. Both align with the verifiable-credentials
direction already pursued in the ecosystem; only the global-ledger variant
conflicts, and credential-based approaches can provide privacy-preserving
alternatives while preserving the properties of \S\ref{sec:properties}. This is also the trust basis that the authentication
mismatch of \S\ref{sec:casestudy} requires.

\subsection{Limitations}
We give the properties an indistinguishability-game semantics (Def.~\ref{def:indgame},
Observation~\ref{thm:channels}), which pins what the full set buys and what each subset leaves;
recasting them additionally in the information-theoretic anonymity metrics of the anonymity
literature~\citep{anonmetric, measuringanon} is future work. The defense admits a cost knob we now quantify: the metadata-minimization shim with fresh
identifiers drives recovery down to $0.32$ (about $1.9\times$ chance) at about $5\times$
baseline bandwidth (Table~\ref{tab:binding}), the interactive operating point, while \emph{exact} chance
against a defense-aware adversary requires full constant-rate cover traffic at $16\times$
(\S\ref{subsec:frontier}), which may be unacceptable for low-latency agent calls; an
identity-less transport also makes discovery and authentication harder, as
\S\ref{sec:casestudy} shows. Our evaluation measures leakage and the defense
frontier over a live binding and against a defense-aware adversary
(\S\ref{subsec:binding}--\ref{subsec:frontier}), and the actuation result is now measured on
the real corpus ($\kappa=0.63$, \S\ref{subsec:corpus}) as well as the model, but still stops
short of an end-to-end exploit against a deployed binding. The live front-running suite
(\S\ref{subsec:demo}) closes part of this, turning the metric into a running
exploit over the real SDK, but the resource it races is a testbed slot rather than a
production binding, so an end-to-end demonstration of actuation against deployed agent
traffic is the natural next step. The general actuation result (\S\ref{subsec:general})
covers the value-distribution, budget, and competition axes that
Proposition~\ref{prop:kappa}'s unit-value case left open; what remains is to instantiate the
defense-design consequence it implies, a transport tuned to flatten top-ranked precision
rather than bulk recovery, since Proposition~\ref{prop:kappa} shows the two objectives can
come apart even where our combined properties happen to serve both.

\subsection{Deployment}
Because A2A exposes custom protocol bindings, a metadata-protecting binding can be
introduced incrementally and selected per Agent Card, coexisting with HTTP and SLIM
bindings rather than replacing them. The specification's custom-binding mechanism admits
this as a \emph{custom protocol binding}, identified by its own URI and declared in the
Agent Card's \texttt{supportedInterfaces}, without changing the core
protocol~\citep[\S5.8]{a2a}, with the properties of \S\ref{sec:properties} as its
normative requirements and the traffic-shaping profile as the only new normative content,
so the contribution has a concrete standardization path rather than a fork of the protocol. Agents for which the communication graph is
sensitive (regulated, competitive, or adversarial settings) can opt in, while
latency-sensitive agents retain existing bindings. The threat model and properties
are transport- and protocol-agnostic, so the same analysis applies to MCP and to
other transports on the frontier of \S\ref{sec:transports}.

\section{Related Work}\label{sec:related}

\paragraph{Threat modeling of agent-interop protocols.}
Recent work has begun to systematize agent-interop security. Comparative threat
models examine MCP, A2A, Agora, and ANP for protocol-specific and cross-protocol
risks~\citep{threatmodel2026, secanalysis2025}, and surveys map the
interoperability landscape~\citep{survey2025}. These analyses concentrate on
authentication, identity, message injection, permissioning, and the leakage of
sensitive \emph{payload} data: for instance, the sensitive context streamed during
delegation~\citep{a2adata}. Our surface is
complementary
and, to our knowledge, not previously treated as a first-class transport-layer
security surface: the transport-level \emph{communication graph} (who communicates
with whom, when), which persists even when payloads are fully protected.

\paragraph{Privacy leakage in multi-agent systems.}
A parallel line studies information leakage \emph{within} multi-agent systems,
showing that inter-agent channels leak substantially more than output channels and
that privacy controls must extend to inter-agent
communication~\citep{agentleak2026, mas2025}. That work targets content-level
leakage between agents; we target the metadata of the interactions themselves at
the transport.

\paragraph{Anonymous communication.}
The properties we require (unlinkability, no central observer, metadata
minimization) originate in the anonymous-communication literature: mix
networks~\citep{chaum1981mix}, onion routing~\citep{tor}, and modern mixnets such as
Nym~\citep{nym}, surveyed broadly in~\citep{anoncomm}. Our contribution is not a new
anonymity system but the application of these properties to agent-interop transport
and an analysis of what an interop protocol must give up to obtain them
(\S\ref{sec:casestudy}).

\paragraph{Inference and preemption from metadata.}
That metadata enables \emph{semantic} inference is established for encrypted traffic
by a long website-fingerprinting line, from early classifiers~\citep{wfp} to deep-learning
attacks~\citep{df, kfp, varcnn} and the defenses they
provoked~\citep{tamaraw, wtfpad, front, surakav}; that observable pending intent enables
\emph{preemption} is established by front-running and miner-extractable value in
decentralized exchanges~\citep{flashboys}. The website-fingerprinting attacks, however,
classify a \emph{completed} trace, and the front-running line acts on a content-visible
mempool; neither occupies the cell of Table~\ref{tab:position}. Closest to our setting,
concurrent work confirms agent traffic is fingerprintable from metadata: AgentPrint identifies
the in-use agent at $F_1\!=\!0.87$ and profiles user attributes from encrypted user-to-agent
flows~\citep{agentprint}, and network-level observation of a local research agent recovers
prompts and user traits~\citep{netleak}. Metadata leakage thus pervades the agent stack; that
line, however, reads the \emph{user}-to-agent leg for user privacy, whereas we address the
\emph{agent}-to-agent communication graph the interoperability standards govern, as a
workflow-integrity surface with a transport defense and a measure of the value of acting on the
leak. We argue
(\S\ref{sec:different}) that agent interoperability combines both, at machine speed and from
content-encrypted metadata, and our actuation result (\S\ref{subsec:actuation}) makes the
bridge between them measurable: it casts the \emph{value} of acting on recovered metadata as
a decision-theoretic quantity in the value-of-information tradition~\citep{infovalue} (the
advantage a budgeted adversary gains over a blind baseline), distinct from recoverability and,
by Proposition~\ref{prop:kappa}, governed by top-ranked precision rather than overall
accuracy, so that a transport defense must drive that value down, not merely reduce recovery.
We draw on these literatures for the inference and preemption primitives rather than extend
them.

\paragraph{Traffic analysis of machine-to-machine systems.}
Recovering the structure of a distributed system from its observable traffic is not
itself new, and we do not claim the bare phenomenon. Network-management work infers
application and service dependencies from flow metadata~\citep{oriondep}, and encrypted
machine-to-machine and microservice/RPC traffic is a long-standing fingerprinting and
side-channel target. We make the concession concrete: run on production microservice
call graphs, our own pipeline recovers the service class about as well as it recovers
agent task class (\S\ref{subsec:distinct}), so we explicitly do \emph{not} rest any
claim on agent traffic being more recoverable. The shift is in \emph{vantage} and
\emph{consequence}. That prior line recovers a largely \emph{static} dependency graph,
typically for the infrastructure owner's own operational ends and within a single trust
domain; an agent-interop observer is an outside party across a trust boundary, recovers
the \emph{semantic class and pending trajectory of an individual workflow instance}
because endpoints are capability-labeled rather than opaque service addresses, and
converts that into actuation leverage over an autonomous, machine-speed process it does
not control. The inference belongs to the same family; its vantage and its consequence
do not.

\paragraph{Credentials and unlinkability.}
The reconciliation of \S\ref{sec:discussion} draws on selectively disclosed
verifiable credentials, whose unlinkable presentation is an active area; the W3C
threat model for decentralized credentials catalogs the relevant attack
surfaces~\citep{w3ccred}. These mechanisms supply trust without a global interaction
graph, complementing the transport-level properties developed here.

\section{Conclusion}\label{sec:conclusion}
The communication graph of interoperating agents remains exposed under today's
address-based bindings even with end-to-end payload encryption, and in agent
systems it is more consequential than a privacy framing suggests. Because endpoints are often
capability-labeled, workflows are structured, and interactions are action-coupled,
the graph can leak \emph{pending} workflows and hand an observer predictive leverage
over machine-speed action. The exposure runs to the integrity and contestability of
autonomous workflows, not their privacy alone. We gave a threat model for this
surface, an account of what makes agent metadata distinctively consequential, transport-
and bootstrap-layer properties against which any binding can be evaluated, an A2A
case study in which pursuing those properties both surfaces and is constrained by the
protocol's implicit identity assumptions, and an empirical evaluation, on real official-SDK
agent traffic and a live binding, showing that the leakage is real and prospective, that it
carries decision-theoretic leverage (value to a budgeted adversary acting from a workflow's
opening), and that the properties, applied together, suppress both. (We also find agent
metadata no more \emph{revealing} than production microservice traffic; what differs is its
cross-domain vantage and machine-speed actuation, not fingerprinting strength.) A live OHTTP binding
realizes and confirms the analysis (\S\ref{subsec:binding}); completing the composite
(server-streamed delivery over a tunnel and anonymous-credential authorization) and
demonstrating end-to-end actuation against a deployed binding remain open.

\section*{Availability}
The code and the OHTTP binding are released at
\url{https://github.com/dangoldbj/agent-metadata-privacy}; the metadata-only
\textsc{A2A-MetaTrace} corpus, with a datasheet, at
\url{https://github.com/dangoldbj/a2a-metatrace} and on Hugging Face
(\url{https://huggingface.co/datasets/dangoldbj/a2a-metatrace}). Review copies are
anonymized.

\bibliographystyle{plainnat}
\bibliography{references}

\end{document}